\documentclass[aps, prb,reprint,10pt,superscriptaddress,amssymb,amsfonts,longbibliography]{revtex4-1}

\usepackage{amsmath}
\usepackage{graphicx}
\usepackage{dcolumn}
\usepackage{bbm}
\usepackage{subfigure}
\usepackage{amssymb}
\usepackage[colorlinks,bookmarks=false,citecolor=blue,linkcolor=red,urlcolor=blue]{hyperref}
\usepackage{wasysym}
\usepackage{MnSymbol}
\usepackage{color}
\usepackage[normalem]{ulem}

\stepcounter{secnumdepth}
\stepcounter{tocdepth}

\def\br{\boldsymbol{r}}

\begin{document}

\title{Fracton topological order from Higgs and partial confinement mechanisms \\ of rank-two gauge theory}

\author{Han Ma}
\affiliation{Department of Physics, University of Colorado, Boulder, Colorado 80309, USA}
\affiliation{Center for Theory of Quantum Matter, University of Colorado, Boulder, Colorado 80309, USA}

\author{Michael Hermele}
\affiliation{Department of Physics, University of Colorado, Boulder, Colorado 80309, USA}
\affiliation{Center for Theory of Quantum Matter, University of Colorado, Boulder, Colorado 80309, USA}

\author{Xie Chen}
\affiliation{Department of Physics and Institute for Quantum Information and Matter, California Institute of Technology, Pasadena, California 91125, USA}

\date{\today}
\begin{abstract}
Fractons are gapped point-like excitations in $d=3$  topological ordered phases whose motion is constrained. They have been discovered in several gapped models but a unifying physical mechanism for generating them is still missing. It has been noticed that in symmetric-tensor ${\rm U}(1)$ gauge theories, charges are fractons and cannot move freely due to, for example, the conservation of not only the  charge but also the  dipole moment. To connect these theories with fully gapped fracton models, we study Higgs and partial confinement mechanisms in rank-2 symmetric-tensor gauge theories, where charges or magnetic excitations, respectively, are condensed. Specifically, we describe two different routes from the rank-2 ${\rm U}(1)$ scalar charge theory to the X-cube fracton topological order, finding that a combination of Higgs and partial confinement mechanisms is necessary to obtain the fully gapped fracton model. On the other hand, the rank-2 $\mathbb{Z}_2$ scalar charge theory, which is obtained from the former theory upon condensing charge-2 matter, is equivalent to four copies of the $d=3$ toric code and does not support fracton excitations. We also explain how the checkerboard fracton model can be viewed as a rank-2 $\mathbb{Z}_2$ gauge theory with two different Gauss' law constraints on different lattice sites.
\end{abstract}

\maketitle

\section{Introduction \label{sec:intro}}

Topological order in three dimensional systems can exhibit completely new features that are not present in lower dimensions. For example, following earlier work of Chamon,\cite{chamon2005quantum,bravyi2011topological} Haah proposed a three dimensional exactly solvable model -- the cubic code -- with topological properties that are very different from all previously known examples.\cite{haah2011local} The cubic code model is topological in the sense that, when defined on a three torus, the model is gapped and has a ground state degeneracy that is robust against any local perturbation. The ground state degeneracy, in sharp contrast to usual three dimensional topological models, changes with system size and is upper bounded by an exponential of the linear system size\cite{haah2011local,bravyi2013quantum}. Similar to usual topological states, the cubic code supports fractional point-like excitations. However, the point-like excitations cannot move freely in three dimensional space. Instead, sets of four point excitations move in a coordinated way, and they can only be separated from each other by a fractal-shaped operator. Similar fractal structures have been found in a class of fractal quantum codes.\cite{haah2011local,yoshida2013exotic,hsieh2017fractons}

Beginning with Chamon, and more recently, a class of gapped three dimensional topological models have been discovered which also host point-like excitations that cannot move freely, but without fractal structure.\cite{chamon2005quantum,vijay2015new,vijay2016fracton,hsieh2017fractons,petrova2017simple, slagle2017fracton} In general, whether or not fractal structure is present, such point-like excitations are dubbed `fractons' or `sub-dimensional particles', and the corresponding models are said to have fracton topological order. The motion of the excitations differs from model to model: some are sub-dimensional particles that move only along a line or in a plane; some are fractons that can only move in coordination with others. This leads to a range of fundamental questions that need to be addressed: What are the universal physical properties characterizing a fracton topological phase? What kinds of fracton topological order are possible? What are physical mechanisms leading to fracton topological order?

Fracton topological order has been studied from a number of different perspectives,\cite{prem2017glassy,slagle2017quantum,slagle2017x,shirley2017fracton,shi2017decipher,ma2017topological,he2017entanglement,schmitz2017recoverable,vijay2017generalization} and several physical mechanisms leading to it have been proposed.
Some fracton phases can be obtained via coupled-layer constructions, where the appearance of fractons is driven by  `particle loop condensation,' starting from a system of decoupled two-dimensional topological states, \cite{vijay2017isotropic,ma2017fracton} obtained by coupling one-dimensional chains.\cite{halasz2017fracton}  Ref.~\onlinecite{hsieh2017fractons} introduced parton theories of fracton states, providing a route to construct variational wave functions in more physically realistic models.

In a closely related development, it was pointed out by Pretko that fractons appear in higher rank ${\rm U}(1)$ gauge theories, \emph{i.e.} those where the electric field and vector potential are symmetric tensors of rank two or higher.\cite{pretko2016subdimensional,pretko2016generalized} In a vector gauge theory, Gauss' law leads to the conservation of total charge; once a positive-negative charge pair is created, each of of the charges can move freely in space without violating charge conservation. For a higher rank gauge theory, the situation can be very different. A modified Gauss's law can lead to the conservation of not only the total charge, but also the conservation of dipole moment, quadrupole moment, etc. Because of the extra conservation laws, the charge excitations cannot move freely any more -- they become fractons (or sub-dimensional particles). We also note several earlier studies of higher-rank ${\rm U}(1)$ gauge theories, although the restricted mobility of charged excitations was not pointed out.\cite{xu2006novel,xu2006gapless,pankov2007resonating,xu2008resonating,xu2010emergent,rasmussen2016stable}

Higher rank ${\rm U}(1)$ gauge theories, being studied extensively recently\cite{prem2018emergent,pretko2017emergent,pretko2017finite,pretko2017fracton,gromov2017fractional}, are different from fracton topological phases as they exhibit gapless photon modes. Is it possible to remove the gapless modes and make a connection with the gapped fracton models mentioned above. A natural way to remove gapless modes in a continuous gauge theory is to `Higgs' the gauge field and reduce the gauge group to a discrete one.\cite{fradkin1979phase} For example, when the normal vector ${\rm U}(1)$ gauge theory is Higgsed down to $Z_2$, we get a $Z_2$ gauge theory -- the ${\rm U}(1)$ gauge charge reduces to the $Z_2$ gauge charge; the ${\rm U}(1)$ flux loop reduces to the $Z_2$ flux loop. Similarly, if we Higgs a gapless higher rank ${\rm U}(1)$ gauge theory, we get a gapped higher rank $Z_N$ gauge theory, which seems to give a natural way to generate gapped fracton topological models. 

Surprisingly, as we show in this paper, the fracton nature of the charge excitations may be lost via the Higgs mechanism. This conclusion was also addressed in a recent study.\cite{slagle2017quantum} Here, we show in detail that while the scalar charge rank-2 ${\rm U}(1)$ gauge theory contains fractons, its Higgsed version does not and is equivalent to several copies of a discrete vector gauge theory. To arrive at a gapped fracton phase via the Higgs mechanism, something more is needed.

\begin{figure}[h]
\includegraphics[width=.45\textwidth]{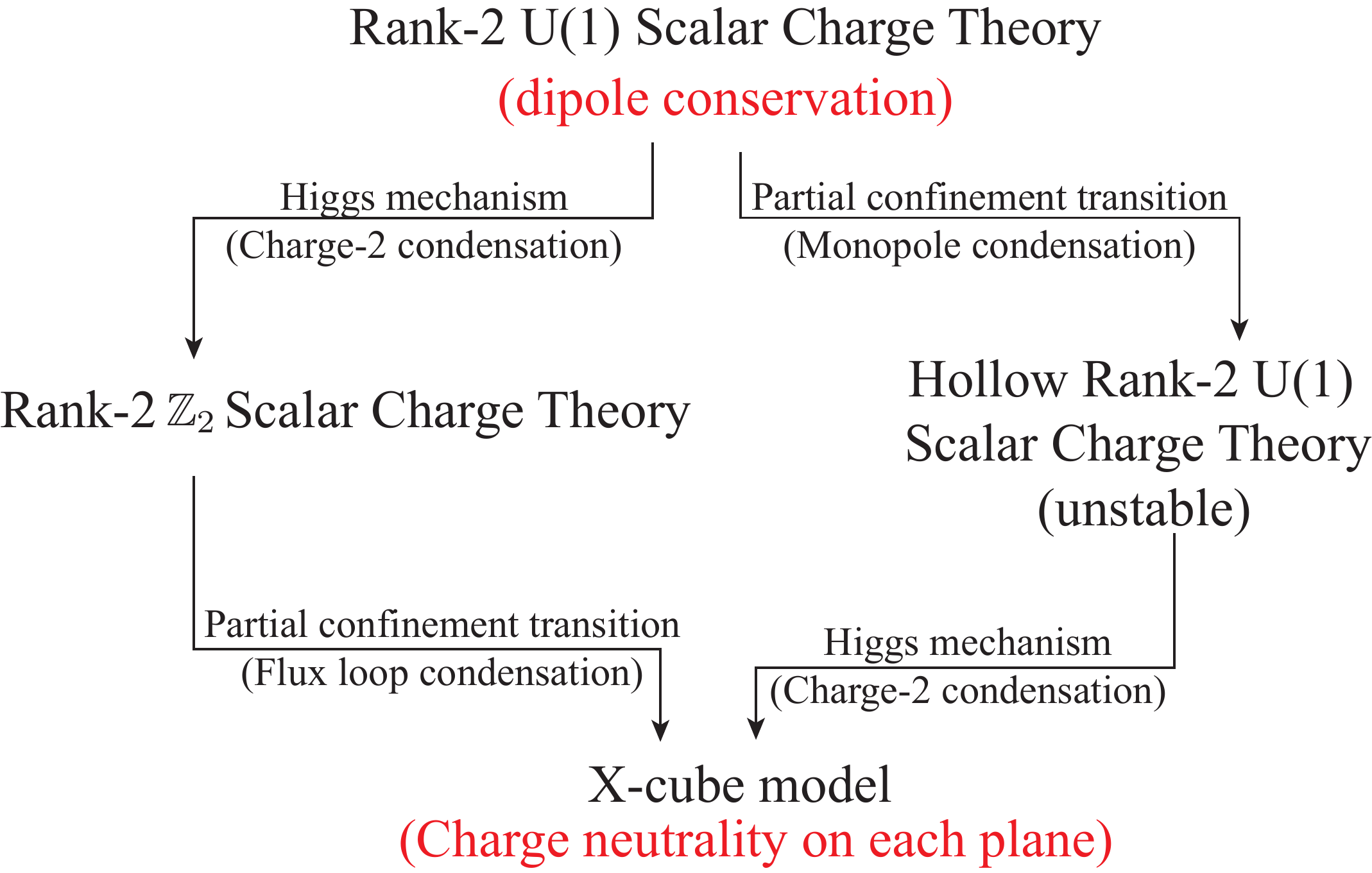}
\caption{Two paths from the rank-2 ${\rm U}(1)$ scalar charge theory to the X-cube model.
\label{fig:higgs}}
\end{figure}

In particular, we discuss two paths that lead from the scalar charge rank-2 ${\rm U}(1)$ gauge theory to the fracton topological order of the X-cube model, a gapped fracton model introduced in Ref.~\onlinecite{vijay2016fracton}.\footnote{The ground state degeneracy on a 3-torus of the exactly solvable X-cube model is stable under local perturbation. This can be verified using degenerate perturbation theory. If we add a local perturbation $V= \lambda\sum \mathcal{O}_{loc}$ to the Hamiltonian, the matrix elements in the resulting effective Hamiltonian for the degenerate ground state space are proportional to $\tilde{\lambda}^L$, where $\tilde{\lambda} \propto \lambda$ is a constant, and $L$ is the linear size of the system.  This holds because only logical operators supported on a region of size $ L$ or larger have non-vanishing matrix elements within the ground state subspace. Since the degenerate subspace has dimension $\sim c^L$ for some constant $c$, the matrix Frobenius norm is bounded by $(c \tilde{\lambda})^L$, which in turn bounds all the eigenvalues. Therefore, as long as $\lambda$ is below some finite threshold, the splitting of the ground state subspace is exponentially small and approaches to zero in the thermodynamic limit. A very similar argument applies to many other gapped fracton models.}  As shown in Fig.\ref{fig:higgs}, in one of the paths we can first Higgs the rank-2 ${\rm U}(1)$ gauge theory by condensing a charge-two matter field, arriving at four copies of $Z_2$ vector gauge theory (the toric code). Then we can condense certain flux loops to partially confine the gauge fields and arrive at the X-cube topological order. 

The second path to the X-cube topological order begins by first condensing certain monopole excitations of the rank-2 ${\rm U}(1)$ gauge theory, to arrive at a distinct `hollow' rank-2 ${\rm U}(1)$ theory, whose field tensors only have the off-diagonal components. This theory was studied previously in Ref.~\onlinecite{xu2008resonating}, where it was shown that it is unstable to confinement arising from proliferation of instantons. Nonetheless, upon condensing charge-two matter in this theory, we again obtain the X-cube fracton topological order.

Similar to the X-cube model, the checkerboard fracton model\cite{vijay2016fracton} can also be interpreted as a rank-2 $\mathbb{Z}_2$ scalar charge theory. An important difference is that two different forms of Gauss' law are alternately enforced on the even/odd layers of the system.

We would like to emphasize that in this paper we are not concerned with the critical properties of transitions between phases, or even whether continuous transitions exist.  Instead, we are interested in physical mechanisms by which one phase can be driven into another phase, by condensation of some excitations. Such mechanisms can be considered independent of critical phenomena, and indeed are still relevant even in cases where a transition is driven first order by fluctuations.  All transitions between phase in this paper are discussed in this spirit, leaving the study of critical phenomena for future work.

The paper is organized as follows. We begin our discussion by reviewing rank-2 ${\rm U}(1)$ gauge theory with scalar charge in Sec.~\ref{sec:rank2_U1}, including a ``hollow'' version of this theory that has only off-diagonal elements in the electric field and gauge field tensors.
Sec.~\ref{sec:higgs_conserve} studies the effect of Higgsing the conservation laws of higher-rank gauge theories, and the resulting predictions for the mobility of charges in the corresponding gapped states. Sec.~\ref{sec:higgs} begins the left-hand branch of Fig.~\ref{fig:higgs}; we couple the rank-2 ${\rm U}(1)$ scalar charge theory to charge-2 matter, and condense the charge-2 matter to obtain the rank-2 $\mathbb{Z}_2$ scalar charge theory. This theory does not support fractons and is found in Sec.~\ref{sec:z_2-model} to be equivalent to four copies of the $d=3$ toric code, with some further details given in Appendix~\ref{app:degeneracy}. We complete the left-hand branch of Fig.~\ref{fig:higgs} in Sec.~\ref{sec:selective}, where we show that condensing flux loops in a selective manner results in the X-cube model. Appendix~\ref{app:rank-2_Z2_2d} studies a two-dimensional version of the rank-2 $\mathbb{Z}_2$ scalar charge theory on the square lattice, showing it is equivalent to three copies of the $d=2$ toric code.

The right-hand branch of Fig.~\ref{fig:higgs} is taken in Sec.~\ref{sec:hollow}. First, in Sec.~\ref{sec:partial-confinement}, we implement an electric-magnetic duality transformation to describe the gapped magnetic excitations of the rank-2 ${\rm U}(1)$ scalar charge theory. These excitations are point-like magnetic monopoles that move in two-dimensional planes, and we show that condensing these excitations in their planes of motion leads to the hollow ${\rm U}(1)$ gauge theory. Sec.~\ref{sec:xcube-from-hollowU1} then shows that condensing charge-2 matter in the hollow ${\rm U}(1)$ gauge theory leads to the X-cube model. Finally, in Sec.~\ref{sec:checkerboard} we show that the checkerboard fracton model can be interpreted as a rank-2 $\mathbb{Z}_2$ gauge theory, with two different forms of Gauss' law on different lattice sites. The paper concludes with a discussion in Sec.~\ref{sec:discussion}.

\section{Review: rank-2 ${\rm U}(1)$ gauge theory \label{sec:rank2_U1}}

Let us start with a brief review of higher rank ${\rm U}(1)$ gauge theories studied in Ref. \onlinecite{pretko2016generalized,pretko2016subdimensional}. Unlike the higher form gauge theories with antisymmetric tensors as their gauge fields, which do not give us new topological orders in three dimensions, the higher rank gauge fields are symmetric tensors.  This difference leads to deconfined phases with sub-dimensional charges in $d \geq 3$.

In this paper, we focus on rank-2 ${\rm U}(1)$ gauge theories and related phases\cite{pretko2016subdimensional,pretko2016generalized}. While we will be interested in studying these theories defined on the lattice, it is useful to first describe them without dynamical matter in the continuum. We note that lattice regularization plays a more important role for these theories as compared to conventional gauge theories, and it is not clear whether these theories make sense with both dynamical matter and the full symmetry of Euclidean three-dimensional space; see Ref.~\onlinecite{fractons_review} for a brief discussion of this issue.

The electric field $E_{\mu \nu}$ and vector potential $A_{\mu \nu}$ are both symmetric tensors, with the Greek indices running over spatial directions, \emph{i.e.} $\mu, \nu = x,y,z$.   Eigenvalues of $E_{\mu \nu}$ and $A_{\mu \nu}$ are real numbers, and we have the commutation relations
\begin{equation}
\left[ A_{\mu \nu} (\br), E_{\lambda \sigma}(\br') \right] = -i ( \delta_{\mu \lambda} \delta_{\nu \sigma} + \delta_{\mu \sigma} \delta_{\nu \lambda} ) \delta(\br - \br')  \text{,}
\end{equation}
where $\br, \br'$ are positions in $d=3$ space.

Such rank-2 gauge theories can be classified according to their Gauss laws and the nature of the corresponding charges. For example, if the Gauss law contracts all the indices of the tensor electric field and gives scalar charge, we have a {\it scalar charge theory}. On the other hand, if the Gauss law gives a charge transforming as a vector, then we have a vector charge theory.

We consider the scalar charge theory, where the Gauss law is $\partial_\mu \partial_\nu E_{\mu\nu} = \rho$, where  repeated indices are summed over. This Gauss law leads to invariance under  gauge transformations $A_{\mu\nu} \rightarrow A_{\mu\nu} + \partial_\mu \partial_\nu f$, where $f$ is an arbitrary function of spatial position.  From $A_{\mu \nu}$ we can construct the gauge-invariant magnetic field tensor $B_{\mu\nu} = \epsilon_{\mu\lambda \sigma} \partial_\lambda A_{\sigma\nu}$, which is traceless but not symmetric, and satisfies $\partial_{\mu} B_{\mu \nu} = 0$ in the non-compact theory.  Deferring until later a discussion of dynamical matter degrees of freedom, the Hamiltonian density is ${\cal H} = \frac{1}{2} E_{\mu \nu} E_{\mu \nu} + \frac{1}{2} B_{\mu \nu} B_{\mu \nu}$.

The unconventional Gauss law leads to a conservation of both electric charge and dipole moment.  Consider some bounded spatial region $V$,  with boundary $\partial V$.  The total charge in $V$ is given by
\begin{equation}
Q = \int_V \rho \, d^3 \br = \int_{\partial V} \partial_{\nu} E_{\mu \nu}  n_{\mu} dS \text{,}
\end{equation}
where $n_{\mu}$ is a unit vector field normal to $\partial V$ and $dS$ is the surface area element.  Because the right-hand side is a boundary term, this implies that it is impossible to locally create electric charges; of course, this is familiar from vector gauge theory.  Here it is also true that the dipole moment $\boldsymbol{d} = \int_V \br \rho \, d^3 \br$ can be written as an integral of the electric field over $\partial V$, so it is also impossible to locally create dipole moments.  One dramatic consequence of this dipole conservation is that single electric charges are immobile, because moving an electric charge changes the dipole moment.

\begin{figure}[t]
\includegraphics[width=.3\textwidth]{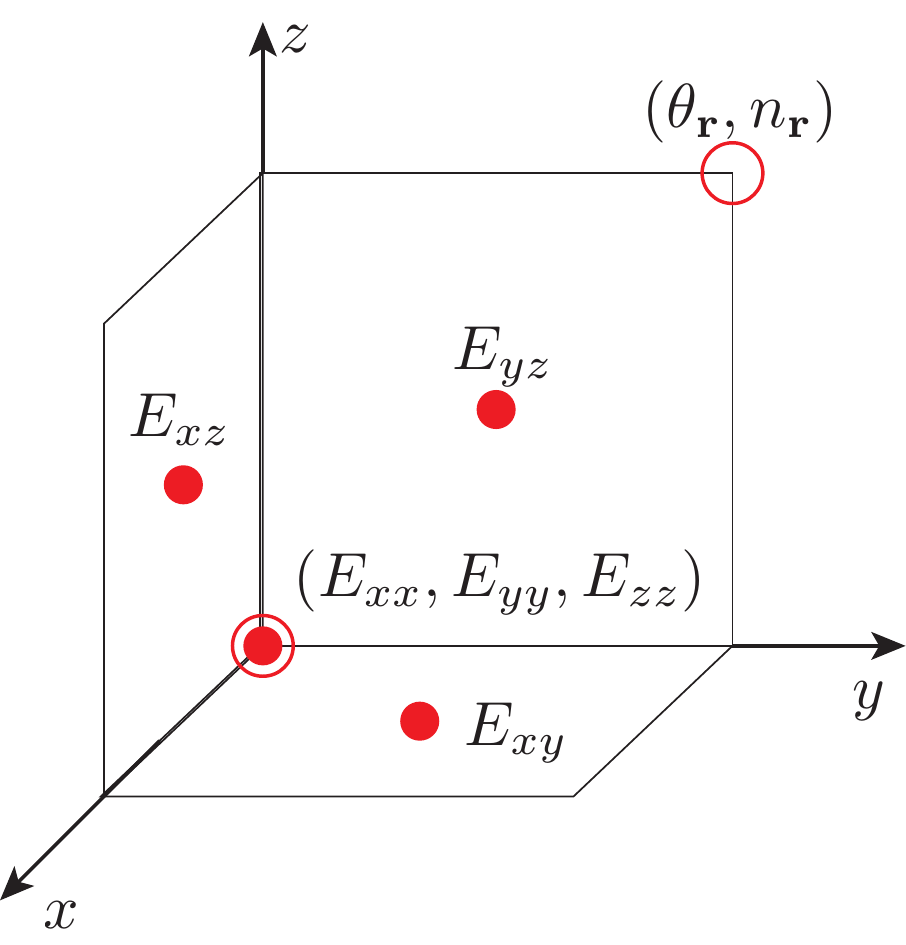}
\caption{Rank-2 ${\rm U}(1)$ gauge theory defined on the cubic lattice. The off-diagonal elements of $E_{\mu \nu}$ and $A_{\mu \nu}$ live on plaquettes, while diagonal elements reside on sites. Gauge charges $n_{\bf r}$, with conjugate  phase $\theta_{\bf r}$, also reside on sites, which are labeled by ${\bf r}$.\label{fig:model_3d}}
\end{figure}

Here, we discuss how to put the rank-2 scalar charge theory on the simple cubic lattice, starting from the continuum theory.  Previous works  considered the related rank-2 vector charge theory, also on the simple cubic lattice \cite{xu2006novel,xu2006gapless,xu2010emergent}.  As shown in Fig.~\ref{fig:model_3d}, the off-diagonal elements $E_{\mu\nu}=E_{\nu\mu}$ (also $A_{\mu\nu}=A_{\nu\mu}$) with $\mu\neq \nu$ are defined on the plaquettes in the $\mu - \nu$ plane, while each site hosts all three diagonal elements $E_{\mu\mu}$ (also $A_{\mu\mu}$). Each conjugate pair $A_{\mu \nu}$, $E_{\mu \nu}$ is an ${\rm O}(2)$ quantum rotor, with $A_{\mu \nu}$ a $2\pi$-periodic phase variable; this makes the theory compact.  The continuum form of the commutation relations implies that $[A_{\mu \nu}, E_{\mu \nu} ] = -i$, with  $\mu \neq \nu$, when the two variables lie on the same plaquette.  For the diagonal components we have $[ A_{\mu \mu}, E_{\mu \mu} ] = - 2i$ (no sum on $\mu$).  This implies that off-diagonal elements of $E$ have integer eigenvalues, while diagonal elements have even integer eigenvalues.  This distinction is somewhat undesirable; we will see how to correct it below.

\begin{figure}[h]
\includegraphics[width=.5\textwidth]{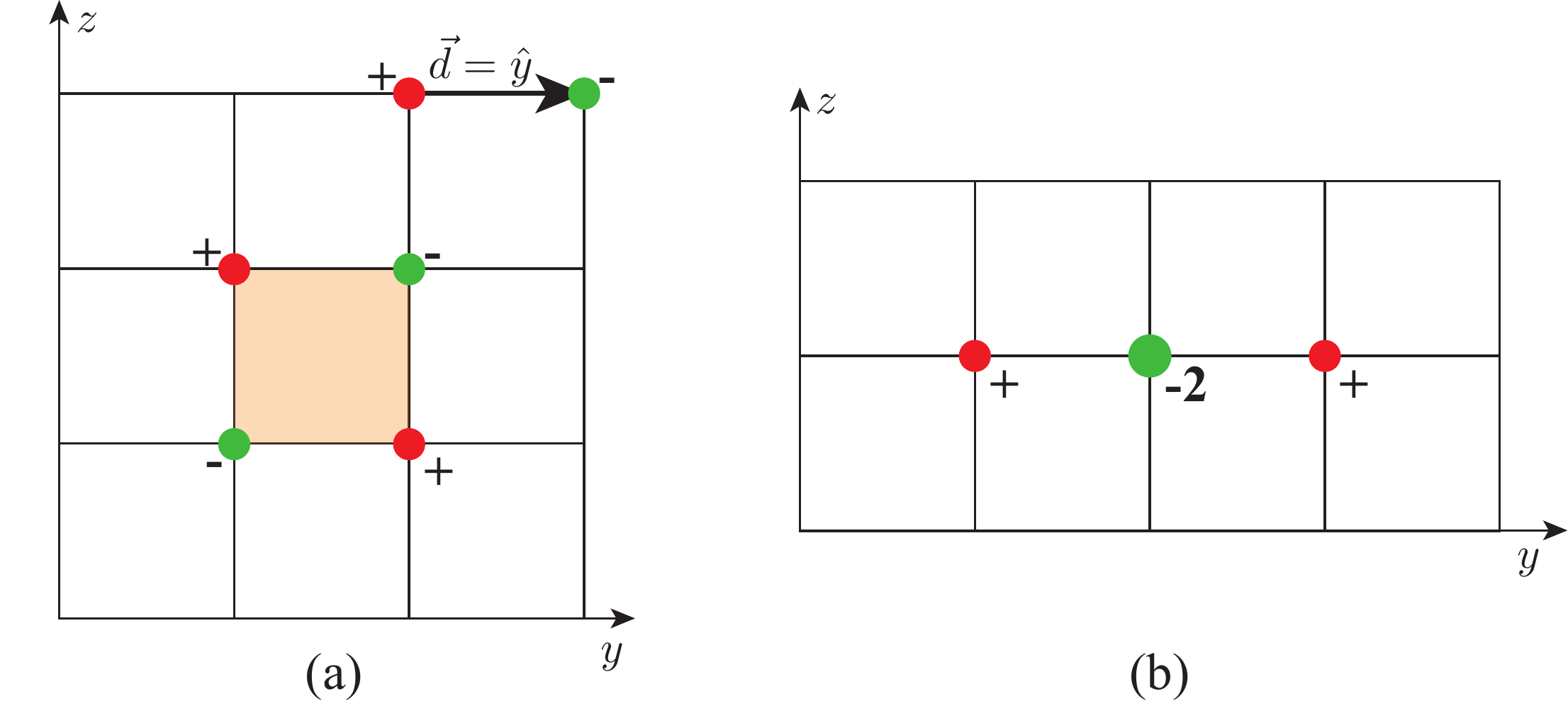}
\caption{Two possible electric charge configurations in the scalar charge theory are shown in (a) and (b). Any configurations related to these by cubic symmetry can also appear. \label{fig:config_scalar}}
\end{figure}

On the lattice, the Gauss law is given by
\begin{equation}
\triangle_\mu \triangle_\nu E_{\mu\nu} = n_{\bf r}\label{eq:Gauss_scalar}
\end{equation}
where $\Delta_{\mu}$ is a finite-difference operator, and $n_{\bf r}$ is the charge at site ${\bf r}$.  In more detail, Gauss law can be written as
\begin{equation}
\begin{aligned}
&2 (\Delta_x  \Delta_y E_{xy} + \Delta_y  \Delta_z E_{yz} +\Delta_x  \Delta_z E_{xz}   ) \\ &+ ( \Delta_x \Delta_x E_{xx} +\Delta_y \Delta_y E_{yy} + \Delta_z \Delta_z E_{zz}) = n_{\bf r} \text{.} 
\end{aligned}
\end{equation}
Here, similar to the difference in commutation relations, there is a factor of two difference in how the diagonal and off-diagonal components of $E$ appear in Gauss' law.  We address these two undesirable features by defining $E'_{\mu \mu} = E_{\mu \mu} / 2$, so that $E'_{\mu \mu}$ takes integer eigenvalues and $[ A_{\mu \mu}, E_{\mu \mu} ] = - i$.  Putting this into Gauss' law, we see that $n_{\bf r}$ is now restricted to be an even integer, so we define the integer-valued charge $n'_{\bf r} = n_{\bf r} / 2$.  Dropping the primes, we then have the commutation relations $[A_{\mu \nu}, E_{\mu \nu} ] = -i$, and the Gauss law
\begin{equation}
\begin{aligned}
& (\Delta_x  \Delta_y E_{xy} + \Delta_y  \Delta_z E_{yz} +\Delta_x  \Delta_z E_{xz}   ) \\ &+ ( \Delta_x \Delta_x E_{xx} +\Delta_y \Delta_y E_{yy} + \Delta_z \Delta_z E_{zz}) = n_{\bf r} \text{.}  \label{eqn:lattice-gauss}
\end{aligned}
\end{equation}
The Gauss' law directly determines those charge configurations that can be created locally, which are illustrated in Fig. (\ref{fig:config_scalar}).  All these ``locally-creatable'' charge configurations have vanishing dipole moment.

We now include dynamical electrically charged matter degrees of freedom, and at the same time describe the lattice Hamiltonian. The matter fields are ${\rm O}(2)$ quantum rotors placed on the cubic lattice sites ${\bf r}$, with number $n_{{\bf r}}$ and phase $\theta_{{\bf r}}$, satisfying $[\theta_{\bf r}, n_{\bf r}] = i \delta_{{\bf r}, {\bf r}'}$.  The Hamiltonian is
\begin{eqnarray}
H &=&  U \sum_{{\bf r},\mu \leq \nu} E^2_{\mu\nu} - K \sum_{{\bf r},\mu,\nu} \cos ( B_{\mu\nu} ) \nonumber \\
&+& u \sum_{\bf r} n^2_{\bf r}  - J \sum_{{\bf r},\mu \leq \nu} \cos \biggl[ \Delta_{\mu} \Delta_{\nu} \theta -A_{\mu \nu} \biggr] \text{.} \label{eqn:h}
\end{eqnarray}
Here the lattice magnetic field is $B_{\mu \nu} = \epsilon_{\mu \lambda \sigma} \Delta_{\lambda} A_{\sigma \nu}$, which can be viewed as a traceless but not symmetric tensor field defined on the dual cubic lattice, where diagonal components reside on dual sites (cube centers of the original lattice), and off-diagonal components reside on dual plaquettes.  The Gauss law is given by Eq.~(\ref{eqn:lattice-gauss}), which leads to gauge transformations
\begin{eqnarray}
A_{\mu \nu} &\to& A_{\mu \nu} + \Delta_{\mu} \Delta_{\nu} f \label{eqn:g1} \\
\theta_{\bf r} &\to& \theta_{\bf r} + f_{\bf r} \text{.} \label{eqn:g2}
\end{eqnarray}
When $u$ and $K$ are the largest energy scales, the charged matter is gapped, and there is a stable deconfined phase that can be described by expanding the $-K \cos(B_{\mu \nu})$ terms to leading (quadratic) order\cite{pretko2016subdimensional}.  This phase has a gapless, linearly dispersing photon mode with five polarizations and gapped charge excitations. There are also gapped magnetic monopole particle excitations, described in Sec.~\ref{sec:partial-confinement}.

We also consider a variant of the rank-2 scalar charge theory, where $E_{\mu \nu}$ and $A_{\mu \nu}$ remain symmetric but have \emph{only} off-diagonal elements.  We refer to this theory as the ``hollow'' rank-2 scalar charge theory, because if we write $E$ and $A$ as $3 \times 3$ matrices, the diagonal elements are zero.  The Gauss law constraint is
\begin{equation}
\Delta_x \Delta_y E_{xy} + \Delta_y \Delta_z E_{yz} + \Delta_{x} \Delta_z E_{x z} = n_{\bf r} \text{,}
\end{equation}
which leads to the same form of gauge transformations as in Eqs.~(\ref{eqn:g1}, \ref{eqn:g2}).  The Gauss law can also be written
\begin{equation}
\frac{1}{2} \Delta_{\mu} \Delta_{\nu} E_{\mu \nu} = n_{\bf r} \text{,}
\end{equation}
where we take the diagonal components of $E$ to be zero.

This Gauss law implies that the charge on every $\{1 0 0\}$ lattice plane is conserved.  Let $R$ be a bounded subset of some lattice plane $p$.   Without loss of generality, we take $p$ to be an $xy$ plane.  We have
\begin{equation}
Q_p = \sum_{{\bf r} \in R} n_{\bf r} = \frac{1}{2} \sum_{{\bf r} \in R} \Delta_{\mu} \Delta_{\nu} E_{\mu \nu} \text{.}
\end{equation}
Each term in the summand contains at least one of $\Delta_x$ or $\Delta_y$, so the sum reduces to a boundary term on $\partial R$, which implies that $Q_p$ is conserved.

If we try to define a magnetic field tensor using the same expression $B_{\mu \nu} = \epsilon_{\mu \lambda \sigma} \Delta_{\lambda} A_{\sigma \nu}$, but setting terms with diagonal components of $A$ to zero, we find that off-diagonal elements of $B$ are no longer gauge-invariant.  Therefore we have only the diagonal elements $B_{\mu \mu}$ (no sum on $\mu$), which satisfy a tracelessness constraint  $\sum_{\mu} B_{\mu \mu} = 0$, where the sum is over the three diagonal elements residing on the same dual lattice site.  This gives us two independent elements of the magnetic field tensor, which is the correct number of degrees of freedom, as there are three independent elements of $A_{\mu \nu}$ and one unphysical gauge degree of freedom.

The Hamiltonian for the hollow gauge theory is
\begin{eqnarray}
H_{{\rm hollow}} &=&  U \sum_{{\bf r},\mu < \nu} E^2_{\mu\nu} - K \sum_{{\bf r}, \mu} \cos ( B_{\mu\mu} ) \nonumber \\
&+& u \sum_{\bf r} n^2_{\bf r}  - J \sum_{{\bf r},\mu < \nu} \cos \biggl[ \Delta_{\mu} \Delta_{\nu} \theta -A_{\mu \nu} \biggr] \text{.} \label{eqn:h_hollow}
\end{eqnarray}
This theory was studied previously in Ref.~\onlinecite{xu2008resonating}, where it was shown that a putative deconfined phase with gapped matter is unstable to confinement via proliferation of instantons.  Even though this theory does not have a deconfined phase (at least with gapped matter), it will still be of interest to us, because it can be Higgsed to obtain the same quantum phase of matter as the X-cube model.

The hollow gauge theory is not just a variant of the ordinary rank-2 scalar charge theory, but it can be obtained from it by a partial confinement transition, as we show in Sec.~\ref{sec:partial-confinement}. Starting from the scalar charge theory, we can make $U$ different for the diagonal and off-diagonal elements of $E$, so that the $E^2$ part of the Hamiltonian becomes
\begin{equation}
H_{{\rm electric}} = U_d \sum_{{\bf r}, \mu} E_{\mu \mu}^2 + U_{od} \sum_{{\bf r}, \mu < \nu} E_{\mu \nu}^2 \text{.}
\end{equation}
Making $U_d$ large results in a state where $E_{\mu \mu} \approx 0$.  This is precisely the hollow gauge theory, which is clear because the Gauss law of the scalar charge theory becomes that of the hollow gauge theory in this limit.  We expect there should be a phase transition as $U_d$ is increased, from the deconfined phase of the scalar charge theory to the (necessarily) confined hollow gauge theory.  Indeed, in Sec.~\ref{sec:partial-confinement}, we show that condensing certain monopoles in the scalar charge theory leads to the hollow gauge theory.

\section{Higgsing the conservation laws \label{sec:higgs_conserve}}

As we mentioned before, the motion of ${\rm U}(1)$ charge is subject to extra conservation laws in  higher rank gauge theories, such as dipole moment conservation. When the ${\rm U}(1)$ gauge theory is Higgsed, we can understand the effects on mobility of charges by directly studying how the conservation laws are modified, which can then be confirmed by a more detailed analysis starting from a Hamiltonian or functional integral formulation.  We thus refer to Higgsing of the conservation laws themselves.  In this section, we study the effect of a condensate of charge-$N$ matter fields, which breaks the gauge group down to $\mathbb{Z}_N$. We discuss charge conservation, dipolar conservation, and the planar conservation law obtained by Higgsing the hollow ${\rm U}(1)$ gauge theory.

Below, all the gauge theories studied have scalar charges living on the cubic lattice sites ${\bf r}=(x,y,z)$, with lattice spacing $a=1$.

\subsection{Charge conservation}

We begin this section with the charge conservation law as a warm up. In a ${\rm U}(1)$ gauge theory, the ${\rm U}(1)$ charge is conserved, meaning, as discussed in Sec.~\ref{sec:rank2_U1}, that the total charge $Q$ in a region cannot be changed by acting with local operators within that region.  Upon Higgsing the theory to $\mathbb{Z}_N$, charge-$N$ objects can appear from and be absorbed into the condensate.  Therefore $Q$ is now well-defined only modulo $N$, and becomes a conserved $\mathbb{Z}_N$ charge.  This of course is familiar from vector gauge theory.  The conservation of $\mathbb{Z}_N$ charge puts no constraints on the mobility of charges.

\subsection{Dipole conservation \label{sec:dipole}}

Now we consider the conservation of dipole moment that arises in the rank-2 scalar charge theory.  On the lattice, the dipole moment $\boldsymbol{d}$ of some region V is given by $\boldsymbol{d} = \sum_{{\bf r} \in V} {\bf r} \, n_{\bf r}$, and is conserved in the sense that it cannot be changed locally.

We now suppose we have a condensate of $N$-charge objects, which results in $\mathbb{Z}_N$ charges. By analogy with charge conservation, we can write down the Higgsed form of dipole conservation, which is simply that $\boldsymbol{d} \mod N$ is conserved, \emph{i.e.} each component of $\boldsymbol{d}$ is separately conserved modulo $N$.  Here it is important that we set the lattice constant to one; then charge $\pm N$ objects appearing from the condensate can change each component of $\boldsymbol{d}$ by integer multiples of $N$.

To understand the effects of the Higgsed dipole conservation law, it is useful first to consider configurations of charges lying on the $x$-axis, \emph{i.e.} with $r_y = r_z = 0$.  We describe such a charge configuration by a set of ordered pairs $\{ (q_1, r_{x1}), (q_2, r_{x2}), \dots \}$, where $q_i$ is the value of the charge at $r_x = r_{xi}$, and the charge is zero for points not listed explicitly.

Dipole conservation tells us that $\{(+1,0), (-1,r_x)\}$ can be locally created only when $r_x$ is a multiple of $N$. Moreover, it is instructive to note that the locally creatable configuration $\{(+1,0),(-1,N)\}$ can  be obtained starting from a charge configuration with vanishing ${\rm U}(1)$ charge and dipole moment, and then exploiting the condensate to add and remove charge-$N$ objects.  We consider the configuration $\{ (+1,0), (-N, i), (N,i+1), (-1, N) \}$, which is easily seen to have vanishing charge and dipole moment.  The charges at site $i$ and $i+1$ can be absorbed into the condensate, and this configuration is thus equivalent to $\{(+1,0),(-1,N)\}$.

On the other hand, the configuration $\{(+1,0),(-1,k)\}$ with $0 < k < N$ has non-vanishing $\boldsymbol{d} \mod N$ and is not locally creatable.  Moreover, a configuration with a $+1$ charge at the origin and a $-1$ charge at ${\bf r} = (r_x, r_y, r_z)$ is easily seen to be have vanishing dipole moment modulo $N$, and thus be locally creatable, only when each of $r_x$, $r_y$ and $r_z$ are integer multiplies of $N$.  From these observations we see that, although the charge in the rank-2 ${\rm U}(1)$ gauge theory is immobile, charges in the Higgsed rank-2 $\mathbb{Z}_N$ gauge theories can hop freely in any direction, but can only hop $N$ lattice spacings at a time along the $x$, $y$ and $z$ directions. This suggests that the rank-2 $\mathbb{Z}_N$ gauge theories obtained in this way are equivalent to the usual vector $\mathbb{Z}_N$ gauge theories. Indeed, we will show in Sec. \ref{sec:z_2-model} that the rank-2 $\mathbb{Z}_2$ scalar charge theory on the cubic lattice is equivalent to four copies of vector $\mathbb{Z}_2$ gauge theories (toric codes). With increasing $N$, the number of distinct particle species is expected to increase. In the limit of $N \rightarrow \infty$, the charge can only hop an infinite number of lattice spacings at a time, and we go back to the ${\rm U}(1)$ case where charge excitations are  immobile.

\subsection{Planar conservation law}
\label{sec:planar}

In the ${\rm U}(1)$ scalar charge theory, the charge and dipole conservation laws impose strong constraints on the motion of excitations. However, after Higgsing, the $\mathbb{Z}_N$ charges turn out to be fully mobile. In order to further restrict the motion of $\mathbb{Z}_N$ charges to be sub-dimensional, we obviously need stronger constraints resulting from more powerful conservation laws.

If the charge on every lattice plane is separately conserved, then clearly single charges will not be able to move.  This is true even for a discrete gauge theory with $\mathbb{Z}_N$ charges, and indeed occurs in the X-cube model (see Sec.~\ref{sec:selective} and Sec.~\ref{sec:xcube-from-hollowU1}).   In other words, this strong conservation law would lead to fractons and sub-dimensional excitations in discrete gauge theory.

In Sec.~\ref{sec:rank2_U1}, we showed that the hollow ${\rm U}(1)$ gauge theory indeed has conservation of charge on each $\{ 100 \}$ lattice plane.  Upon Higgsing this conservation law by condensing charge-$N$ objects, we obtain $\mathbb{Z}_N$ charges that are conserved separately in each plane, which are thus immobile fractons.  At $N=2$, this is exactly the same conservation law as in the X-cube model, and we expect that condensing charge-2 objects in the hollow ${\rm U}(1)$ gauge theory will lead to the X-cube topological order.  This expectation is verified in Sec.~\ref{sec:xcube-from-hollowU1}.

\section{X-cube model via $\mathbb{Z}_2$ scalar charge theory \label{sec:higgs}}

\subsection{Charge-2 condensation \label{sec:charge-2}}

The results on the mobility of $\mathbb{Z}_N$ charges predicted above by Higgsing the conservation laws can be obtained explicitly  by coupling the lattice model  to a charge-$N$ matter field.  We do this here for the rank-2 ${\rm U}(1)$ scalar charge theory.  Starting with the Hamiltonian in Eq.~\ref{eqn:h}, we add a charge-2 matter field on sites ${\bf r}$ with number $N_{\bf r}$ and phase $\Theta_{\bf r}$. We add the following terms to the Hamiltonian in Eq.~\ref{eqn:h}
\begin{equation}
\begin{aligned}
H_{2e} &= u_2 \sum_{\bf r} N_{\bf r}^2 - \Delta \sum_{{\bf r}} \cos \left[ \Theta_{\bf r} -2 \theta_{\bf r} \right] \\
& - J_2 \sum_{{\bf r},\mu \leq \nu} \cos \biggl[ \Delta_{\mu} \Delta_{\nu} \Theta
-2 A_{\mu\nu}({\bf r})\biggr] \text{.}
\end{aligned}
\end{equation}
The Gauss law in Eq.~(\ref{eqn:lattice-gauss}) is modified to become
\begin{equation}
(\Delta_x  \Delta_y E_{xy} + \cdots )  + ( \Delta_x \Delta_x E_{xx} + \cdots) = n_{\bf r} + 2 N_{\bf r} \text{.} 
\end{equation}

Tuning $J_2$ to be large results in a condensation of the charge-2 field $e^{i\Theta}$.  We also take $\Delta$ large for convenience, and we treat the $\Delta$ and $J_2$ cosine terms as constraints.  To solve the constraint imposed by $\Delta \to \infty$, we  define a $\mathbb{Z}_2$ charge creation operator $\tau^z_{\bf r}  \equiv e^{i \zeta_{\bf r}}$, where
\begin{equation}
\zeta_{\bf r} = \frac{1}{2}\Theta_{\bf r} - \theta_{\bf r} =0,\pi \text{.}
\end{equation}
The operator $\tau^z_{\bf r}$ anticommutes with  $\tau^x_{\bf r} \equiv \exp \left[ i \pi n_{\bf r} \right]$, which justifies the notation suggestive of Pauli matrices.

\begin{figure}[h]
\includegraphics[width=.5\textwidth]{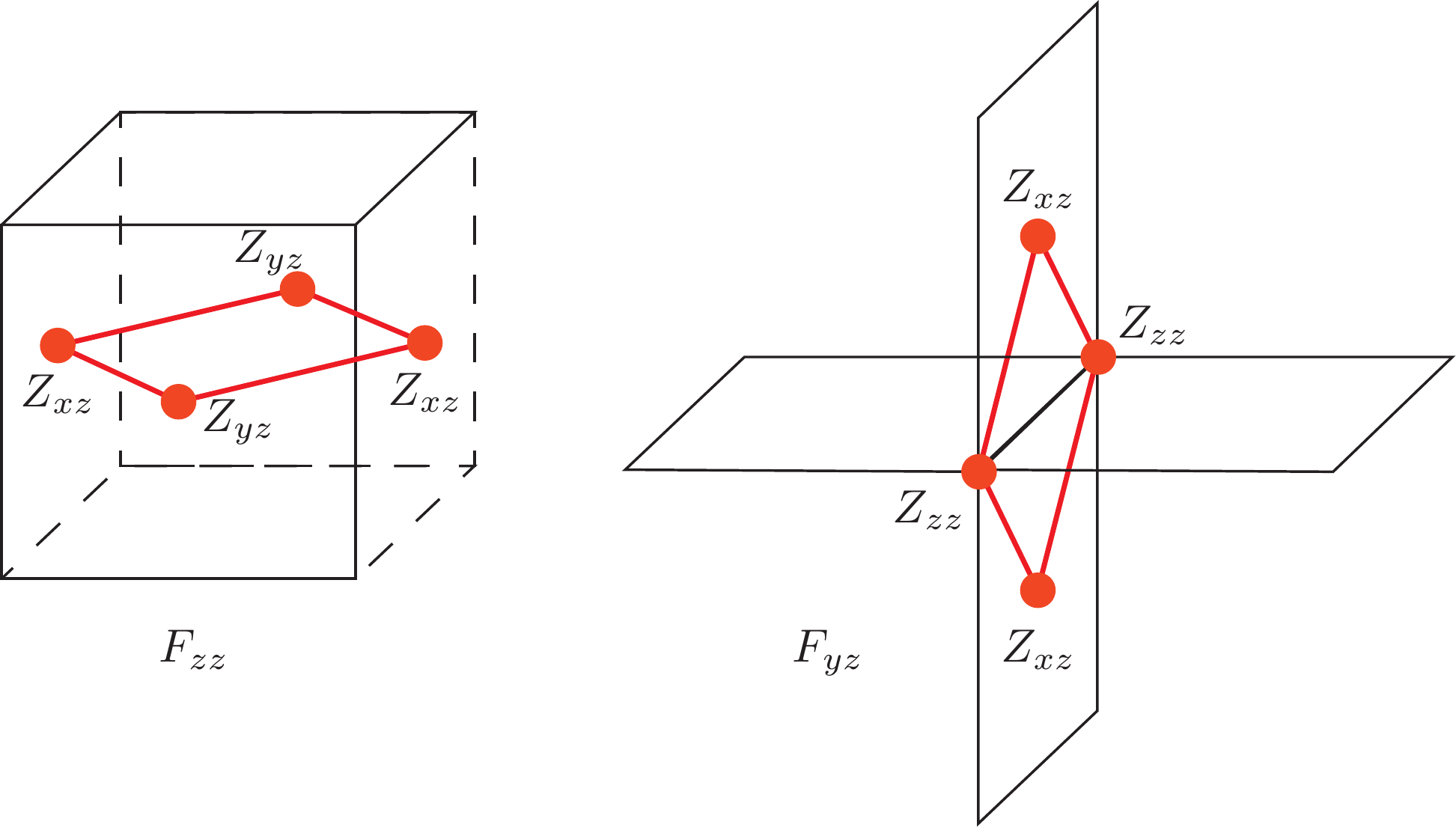}
\caption{Two elements of the ${\mathbb{Z}}_2$ magnetic flux tensor $F_{\mu \nu}$ are shown as products of   $\mathbb{Z}_2$ gauge field operators $Z_{\mu\nu}$. \label{fig:B_element}}
\end{figure}

The second cosine leads to a rank-2 $\mathbb{Z}_2$ gauge field $Z_{\mu\nu} \equiv \exp (i \eta_{\mu\nu})$ on plaquettes ($\mu \neq \nu$) and sites ($\mu=\nu$), where
\begin{equation}
\eta_{\mu\nu} = \frac{1}{2} \Delta_{\mu} \Delta_{\nu} \Theta - A_{\mu\nu}({\bf r})=0,\pi \text{.}
\end{equation}
The  operator $Z_{\mu\nu}$ anticommutes with the rank-2 $\mathbb{Z}_2$ electric field $X_{\mu\nu} \equiv \exp (i \pi E_{\mu\nu})$.  The ${\mathbb Z}_2$ magnetic flux is given by $F_{\mu \nu} = \exp(i B_{\mu \nu} ) = \exp(i \epsilon_{\mu \lambda \sigma} \Delta_{\lambda} \eta_{\lambda \nu} )$. Two elements of $F_{\mu \nu}$ are shown in Fig. (\ref{fig:B_element}). 

We can now write down the effective Hamiltonian describing the system upon condensing $e^{i\Theta}$, in terms of the new  $\mathbb{Z}_2$ variables, which is a rank-2 ${\mathbb{Z}}_2$ scalar charge gauge theory. We have
\begin{equation}
\begin{aligned}
H_{\mathbb{Z}_2} &= -J \sum_{{\bf r},\mu \leq \nu} \exp(i \Delta_{\mu} \Delta_{\nu} \zeta ) Z_{\mu\nu}   \\
&- K \sum_{{\bf r}, \mu,\nu} F_{\mu\nu}
- u \sum_{\bf r} \tau^x_{\bf r} - U \sum_{{\bf r},\mu \leq \nu} X_{\mu\nu} . \label{eq:ham_rank-2_z2}
\end{aligned}
\end{equation}
Here we have replaced the $n_{\bf r}^2$ and $E_{\mu \nu}^2$ terms with $-\tau^x_{\bf r}$ and $- X_{\mu \nu}$, respectively.  These are the simplest terms in the $\mathbb{Z}_2$ variables that penalize non-zero charge and electric field, respectively.

\begin{figure}[h]
\includegraphics[width=.4\textwidth]{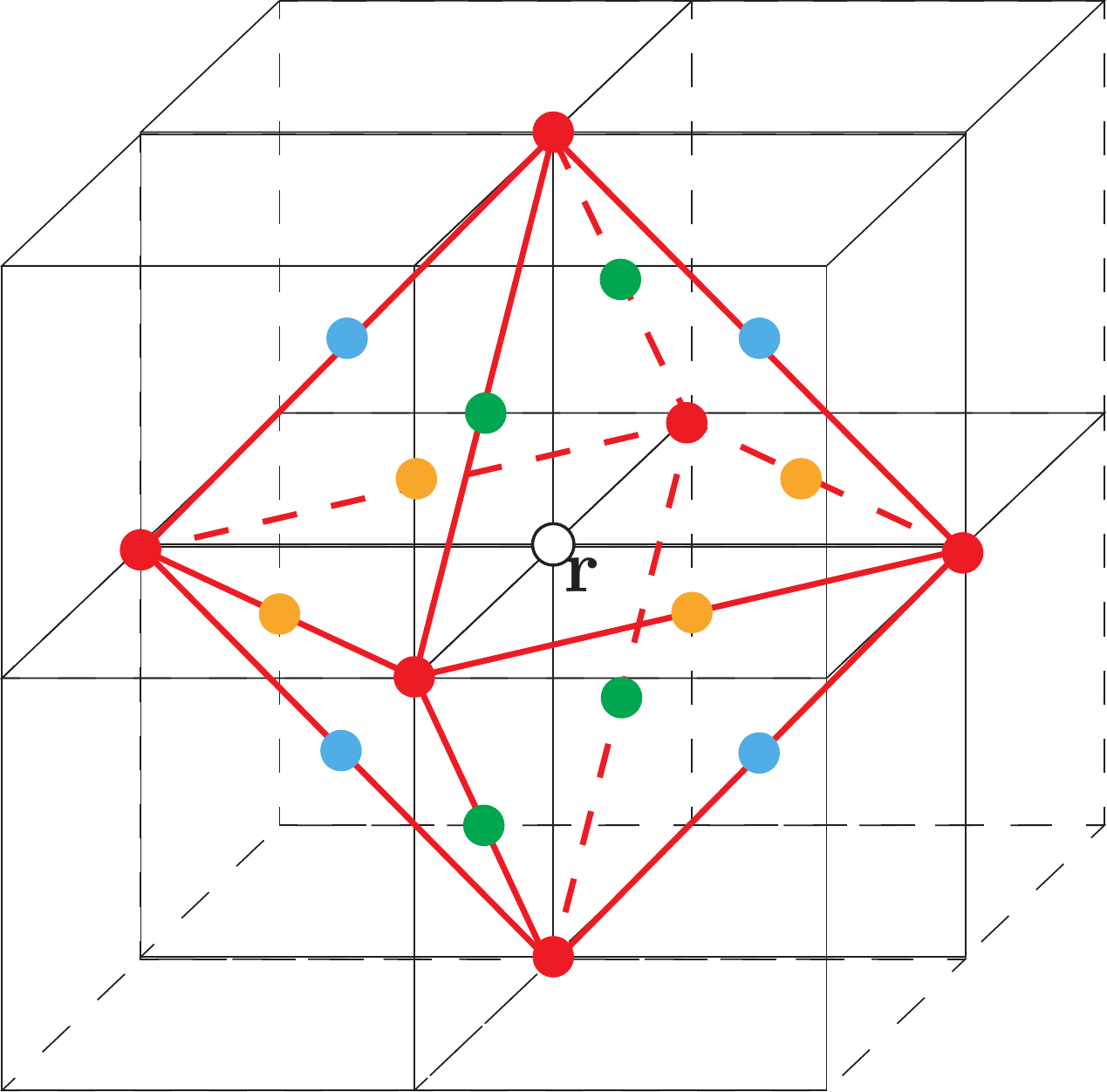}
\caption{The octahedron term $G_{\bf r}$ at ${\bf r}$ involves eighteen Pauli operators denoted as solid dots. Six of them are at sites (red). The other ones are at plaquettes adjacent to ${\bf r}$ in $xy$ (yellow), $yz$ (blue) and $xz$ (green) planes. The open circle shows the location of the charge created by violating this $G_{\bf r}$ term.  \label{fig:term_Gauss}}
\end{figure}

The Gauss' law is obtained by exponentiating that of the ${\rm U}(1)$ theory, and is
\begin{equation}
\exp\Big( i \pi ( \Delta_{x} \Delta_{x} E_{x x} + \cdots ) + (\Delta_{x} \Delta_y E_{xy} + \cdots) \Big) = \tau^x_{\bf r} \text{.}
\end{equation}
The left-hand side, which we denote $G_{\bf r}$,  is a product of 18 $X_{\mu \nu}$ operators on the edges and vertices of an octahedron, as illustrated in Fig.~\ref{fig:term_Gauss}.

We are interested in the deconfined phase of this theory, where magnetic flux is suppressed and where the matter fields are gapped.  An extreme limit of the deconfined phase obtains for $J = U = 0$, where the model becomes exactly solvable.  The $U$ term gives a tension to non-trivial electric field configurations, and, equivalently, leads to fluctuations of the magnetic field.  For large enough $U$, we expect the theory to become confining.  The $J$ term is a kinetic energy for the $\mathbb{Z}_2$ charges, which we expect to condense for large enough $J$.

We will analyze the deconfined phase at its exactly solvable point.  To do this, it will be convenient to exploit a mapping to a local bosonic model with a tensor product Hilbert space (\emph{i.e.}, not a gauge theory).  This mapping works in the same way as the familiar mapping between  vector $\mathbb{Z}_2$ gauge theory and the $\mathbb{Z}_2$ toric code.\cite{kitaev2003fault}  The degrees of freedom in the bosonic model are Pauli operators $\tilde{Z}_{\mu \nu}$ and $\tilde{X}_{\mu \nu}$.  These are related to the gauge theory degrees of freedom by
\begin{eqnarray}
\tilde{X}_{\mu \nu} &=& X_{\mu \nu} \\
\tilde{Z}_{\mu \nu} &=& \exp(i \Delta_{\mu} \Delta_{\nu} \zeta ) Z_{\mu\nu} \text{.}
\end{eqnarray}
Using Gauss' law to express $\tau^x_{\bf r}$ in terms of $G_{\bf r}$, the Hamiltonian for the bosonic model is
\begin{eqnarray}
\tilde{H}_{\mathbb{Z}_2} &=& - K \sum_{{\bf r}, \mu,\nu} \tilde{F}_{\mu\nu}  - u \sum_{\bf r} \tilde{G}_{\bf r} \label{eq:ham_rank-2_z2-bosonic} \\
&-& J \sum_{{\bf r},\mu \leq \nu}  \tilde{Z}_{\mu\nu}  -  U \sum_{{\bf r},\mu \leq \nu} \tilde{X}_{\mu\nu}  \nonumber \text{,}
\end{eqnarray}
where $\tilde{F}_{\mu \nu}$ and $\tilde{G}_{\bf r}$ are products of $\tilde{Z}_{\mu \nu}$ and $\tilde{X}_{\mu \nu}$, respectively, given by the same expressions as $F_{\mu \nu}$ and $G_{\bf r}$.

\subsection{Lattice $\mathbb{Z}_2$ scalar charge theory \label{sec:z_2-model}}

We now study the deconfined phase of the $\mathbb{Z}_2$ scalar charge theory at its exactly solvable point.  We work with the local bosonic model of Eq.~(\ref{eq:ham_rank-2_z2-bosonic}), set $U = J = 0$, and drop the tildes that distinguish between operators acting in the gauge theory and tensor product Hilbert spaces. We will show that this model is  equivalent to four copies of the three dimensional toric code (equivalently, four copies of  $\mathbb{Z}_2$ vector gauge theory).  In the next section, we start from this model and describe how to obtain the X-cube model via a confinement transition.

First, in Appendix~\ref{app:degeneracy}, we establish that the ground state degeneracy ${\rm GSD}$ on a $L \times L \times L$ torus is a constant: $\log_2\left(\textrm{GSD}\right)=12$, independent of system size.  Moreover, the degenerate ground states cannot be locally distinguished, and the model is topologically ordered.  This is consistent with the model being equivalent to four copies of the $d=3$ toric code.

Here, we describe the excited states of the model, which can be labeled by the eigenvalues of the commuting operators $G_{\bf r}$ and $F_{\mu \nu}$.  First, we consider ``electric'' excitations where $G_{\bf r} = -1$ for some sites ${\bf r}$ contained in a bounded region, and where $F_{\mu \nu} = 1$.  We will see that any such excitation can be built up from four independent types of fully mobile ${\mathbb Z}_2$ point charges.

Acting with a $Z_{\mu\nu}$ operator on a ground state creates point-like charges at sites by flipping the sign of some $G_{\bf r}$ eigenvalues.  Depending on whether $\mu = \nu$ or $\mu \neq \nu$, two charge configurations are possible, and are the same as in Fig.~\ref{fig:config_scalar}, but with $\mathbb{Z}_2$ charges. In the linear charge configuration of Fig.~\ref{fig:config_scalar}b, obtained by acting with $Z_{\mu \mu}$, there are now only two nontrivial charges, as the middle one vanishes into the condensate.  Because of this, single charges are able to hop from site to site by an even number of lattice spacings. Therefore, by acting repeatedly with $Z_{\mu \mu}$ operators, we can transform a general electric excitation into one supported on a cube of eight sites ${\bf r} = (r_x, r_y, r_z)$, where $r_x, r_y, r_z = 0,1$, and where we take the origin to be one corner of the cube.

\begin{figure}[h]
\includegraphics[width=.35\textwidth]{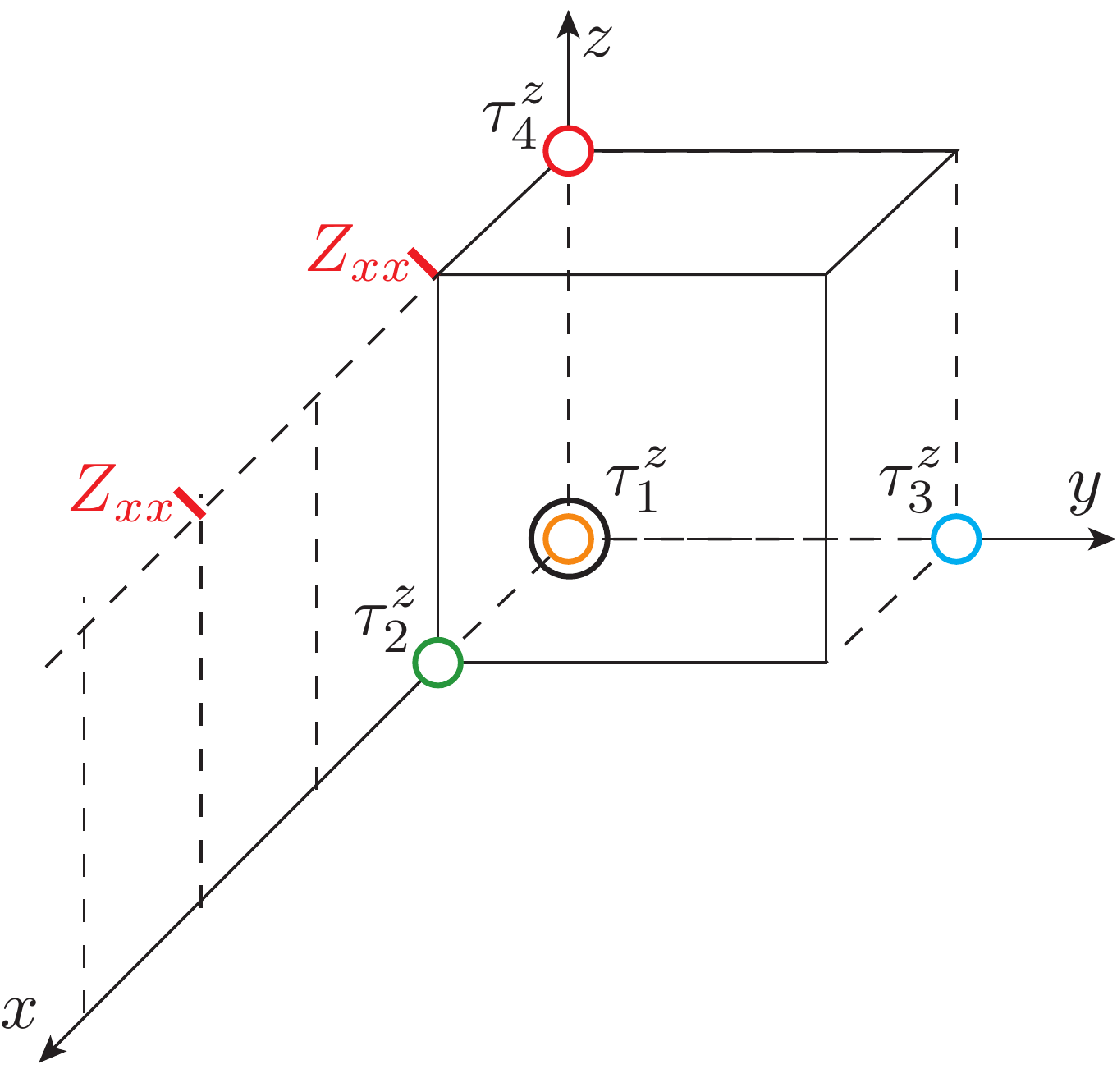}
\caption{Four types of $\mathbb{Z}_2$ gauge charges $\tau^{z}_{1,2,3,4}$ live on different lattice sites as shown. Charges on other sites are equivalent to one of these four types. The large open black circle is a choice of origin, and by definition charges of type $\tau^z_1$ reside at the origin. These charges can be transported by string operators and are freely mobile. A string operator creating the $\tau^z_4$ charge is shown; this operator is a product of $Z_{xx}$ over every other site on a line in the $x$-direction. \label{fig:charges} }
\end{figure}

At this point it might appear that there are eight types of independent charges, corresponding to the eight vertices of the cube.  However, acting with off-diagonal $Z_{\mu \nu}$ operators on the faces of the cube, any charge configuration can be brought to one with excitations at only four of the eight vertices, as shown in Fig.~\ref{fig:charges}.  There are thus four types of $\mathbb{Z}_2$ charges, corresponding to $G_{\bf r} = -1$ at only a single one of the four vertices of Fig.~\ref{fig:charges}.  We label the charge types by $\tau^{z}_{1,2,3,4}$, as shown in the figure.

\begin{figure*}
\includegraphics[width=1.\textwidth]{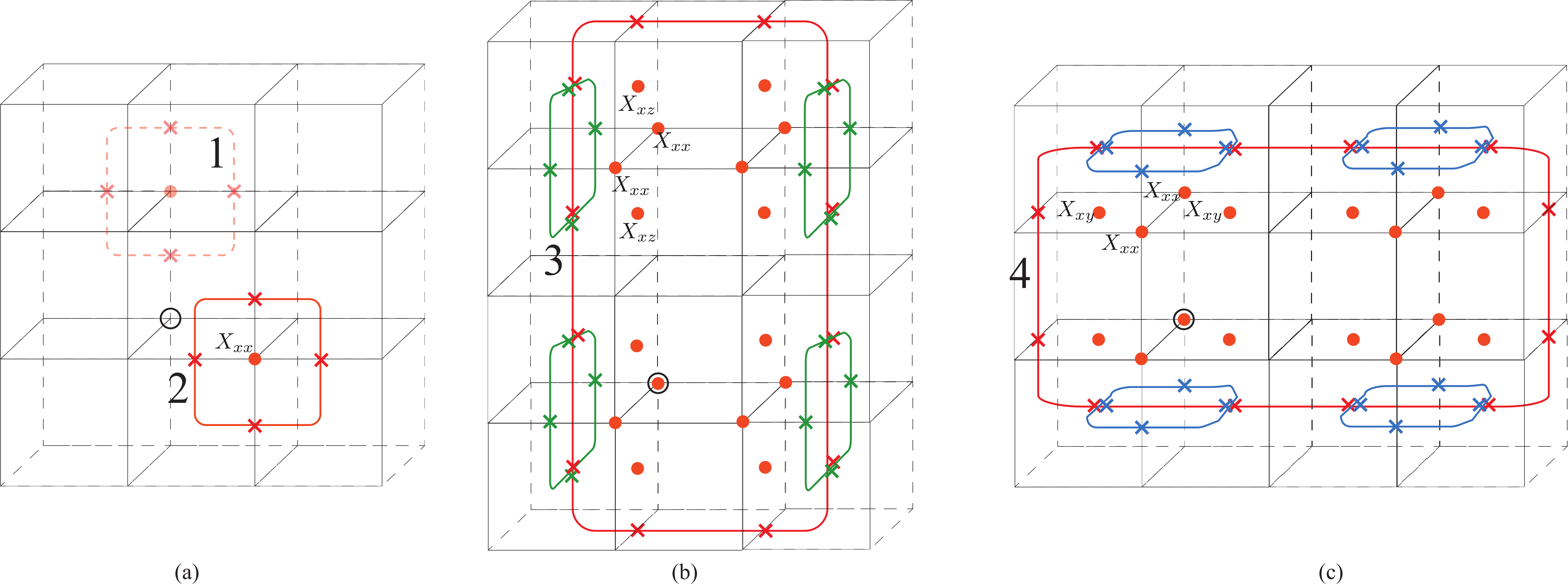}
\caption{Four types of $yz$-plane loop excitations in the rank-2 $\mathbb{Z}_2$ scalar charge theory. In each panel, red dots show the locations of the $X_{\mu \nu}$ operators creating the loop, and the black open circle shows the coordinate origin defined in Fig.~\ref{fig:charges}. (a) shows loops of type 1 and 2, created by acting with $X_{xx}$ within different $yz$-plane layers with even and odd $x$ coordinate. (b) and (c) show two different loop excitations created by acting with a product of $X_{xx}$ together with $X_{xz}$ or $X_{xy}$, respectively. These four loops are distinct excitations because they have different statistical interactions with the four different types of $\mathbb{Z}_2$ gauge charge.
\label{fig:magnetic_excitation}}
\end{figure*}

The existence of four types of $\mathbb{Z}_2$ charge is consistent with equivalence to four copies of the $d=3$ toric code.  To go further, we should also establish the existence of magnetic loop excitations that have non-trivial braiding with the charges.  For a single copy of the $d=3$ toric code, there is a statistical phase $\theta = \pi$ when a point charge braids around a loop.  Here, upon braiding any composite of the elementary charges $\tau^z_{1,2,3,4}$, we expect a statistical phase of $\theta = 0$ or $\theta = \pi$.  We need to  show that there exist four different types of magnetic loops, such that if we braid a single point-like excitation with each of the four loops, the pattern of statistical phases uniquely determines the charge type.   A set of such loop excitations is shown in Fig.~\ref{fig:magnetic_excitation}; in each case the loop and the membrane operator creating it lies in a $yz$ plane.  The statistics of each of the four elementary charges with a given loop is then easily determined by noting whether the string operator transporting the charge in the $x$-direction, which is a product of $Z_{xx}$ operators, commutes or anti-commutes with the loop's membrane operator.  This information is shown in Table~\ref{tab:point-loop}, and it is straightforward to verify that the charge type is fully resolved by braiding with the four different loops.

\begin{table}
\begin{tabular}{c|c}
Loop type & Elementary charges with $\theta = \pi$ statistics \\
\hline
1 &  $\tau^z_2$  \\
\hline
2 & $\tau^z_1, \tau^z_3, \tau^z_4$\\
\hline
3 & $\tau^z_1, \tau^z_2, \tau^z_3$ \\
\hline
4 & $\tau^z_1, \tau^z_2, \tau^z_4$ \\
\hline
\end{tabular}
\caption{In this table, for each type of magnetic loop as shown in Fig.~\ref{fig:magnetic_excitation}, we list the elementary charges $\tau^z_{1,2,3,4}$ that acquire a statistical phase $\theta = \pi$ when braided around the loop indicated.  The statistical phase is $\theta = 0$ for elementary charges not listed.
 \label{tab:point-loop}}
\end{table}

Based on the ground state degeneracy and properties of excitations and logical operators, we conclude that the  $\mathbb{Z}_2$ scalar charge model has the same topological order as four copies of the $d=3$ toric code.  There are no sub-dimensional particle excitations in this system, which is consistent with our prediction from Higgsing the conservation law.

We note that a similar analysis can be carried out in two dimensions, starting with the rank-2 ${\rm U}(1)$ scalar charge theory on the square lattice, and Higgsing it by coupling to charge-2 matter.  There, the resulting $\mathbb{Z}_2$ scalar charge theory is equivalent to three copies of the $d=2$ toric code.  This analysis is presented in Appendix \ref{app:rank-2_Z2_2d}.

\subsection{X-cube model from selective flux loop condensation \label{sec:selective}} 

Here we will describe how to get the X-cube fracton topological order from the rank-2 $\mathbb{Z}_2$ scalar charge theory, by condensing certain flux loops. Before doing this, we discuss a simpler but similar transition from the cubic-lattice $d=3$ $\mathbb{Z}_2$ toric code to a stack of decoupled $d=2$ toric code layers.

\begin{figure}[h]
\includegraphics[width=.4\textwidth]{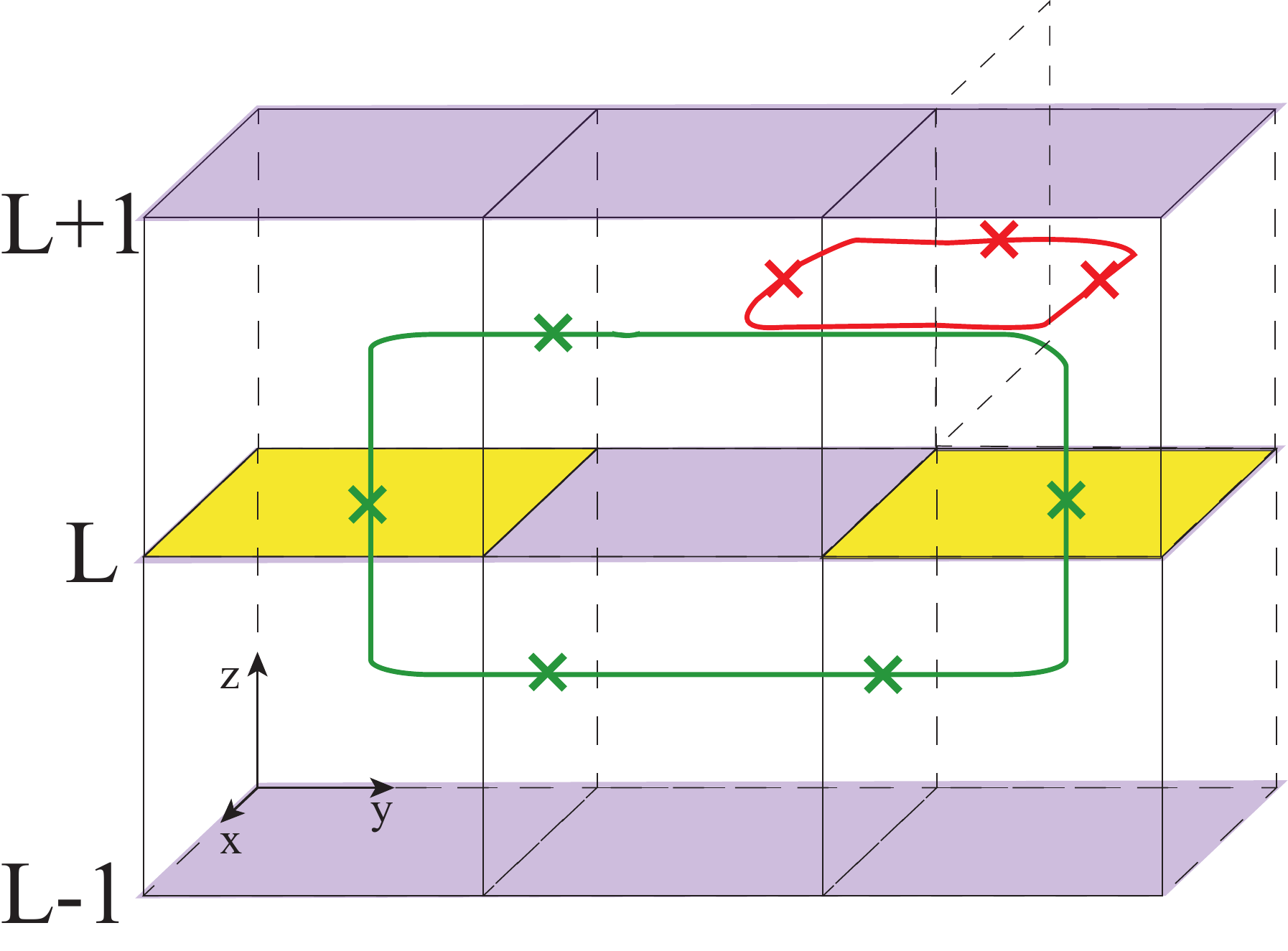}
\caption{Loop condensation (red) between $z = L$ and $z = L\pm1$ layers in the $d=3$ toric code.  The horizontal segments of the green loop disappear into the condensate, while the intersection points with the $z = L$ layer (yellow plaquettes) become gapped point-like excitations that can move freely within the $z = L$ plane. \label{fig:tc_dim_reduction} }
\end{figure}

We consider a perturbed $d=3$ $\mathbb{Z}_2$ toric code, which has a single qubit on each nearest-neighbor link $\ell$ of the simple cubic lattice.  
The Hamiltonian is
\begin{equation}
H_{3dTC} = - \sum_p B_p - \sum_{\bf r} A_{\bf r} - h \sum_{\ell \parallel z} X_\ell
\end{equation}
where $p$ labels square plaquettes, ${\bf r}$ labels cubic lattice sites, and we take $h \geq 0$.  The sum in the last term is over links parallel to the $z$-axis.  The plaquette term is given by  $B_p = \prod_{\ell \in p} Z_\ell$, while the vertex term is $A_{\bf r} = \prod_{\ell \sim {\bf r}} X_\ell$, where latter product is over the six links touching the site ${\bf r}$.  When $h = 0$, we have the usual exactly solvable $d=3$ toric code, which has point charge excitations where $A_{\bf r} = -1$, and flux excitations that are loops along which $B_p = -1$.

Now we consider the effect of the perturbation, which tends to freeze the degrees of freedom between $xy$ plane layers.  In the strong coupling $h \to \infty$ limit, we set $X_\ell = 1$ for all $\ell \parallel z$.  The remaining terms can be treated in first-order degenerate perturbation theory.  Projecting the Hamiltonian into the degenerate subspace, each  $A_{\bf r}$ term becomes a product of $X_\ell$ over the four touching ${\bf r}$ and lying in an $xy$ plane.  Plaquette terms $B_p$ where $p$ lies in an $xy$ plane survive the projection unchanged, while other plaquette terms have vanishing projection into the degenerate subspace.  The resulting Hamiltonian is simply that of a stack of decoupled $d=2$ toric code layers.

The transition from $d=3$ toric code to $d=2$ toric code layers can be interpreted in terms of a condensation of certain flux loops.  Considering first small $h$, we see that acting with $X_\ell$ for $\ell \parallel z$ creates a small flux loop lying within an $xy$ plane between two square lattice layers.  For intermediate $h$, many such $xy$-plane flux loops are created, and eventually we expect these loops will condense at infinite $h$.  The point charge excitations of the $d=3$ toric code can no longer move in the $z$-direction, because they have a statistical interaction with the loop condensate.  However, the charges can still move freely within $xy$-plane square lattice layers, where the loop condensate does not penetrate.  In addition, a flux loop that intersects a single square lattice layer at two points, as shown in Fig.~\ref{fig:tc_dim_reduction}, reduces to two point-like flux excitations at the intersection points, in the presence of the condensate.

Now we return to the rank-2 $\mathbb{Z}_2$ scalar charge theory discussed in the previous section, from which we obtain the X-cube topological order via a similar selective flux loop condensation.  We add the following term to the Hamiltonian
\begin{equation}
H' = - h\sum_{{\bf r}} \sum_{\mu = x,y,z} X_{\mu\mu}({\bf r}) \text{,}
\end{equation}
where we recall that the diagonal components $X_{\mu \mu}$ reside at lattice sites ${\bf r}$.
In the strong coupling limit, $h \to \infty$, $X_{\mu\mu} ({\bf r}) $ is restricted to be $1$.  As in the $d=3$ toric code case, we treat the remaining terms in first-order degenerate perturbation theory.  Projecting the $G_{\bf r}$ term into the degenerate subspace with the projector $P$ results in  
\begin{equation}
P \tilde{G}_{\bf r} P = \prod_{p \sim {\bf r}} X_{\mu\nu}(p) \text{,}
\end{equation}
where the product is over the 12 plaquettes $p$ with a corner at the site ${\bf r}$.  The off-diagonal $F_{\mu \nu}$ terms have vanishing projection into the degenerate subspace, while the diagonal terms $F_{\mu \mu}$ survive unchanged.  The resulting model has only off-diagonal ($\mu \neq \nu$) Pauli operators $X_{\mu \nu}$ and $Z_{\mu \nu}$ residing on plaquettes.  Replacing the cubic lattice with its dual, the plaquette variables become link variables $X_\ell$ and $Z_\ell$, where $\ell$ labels dual lattice links.  It is straightforward to see that 
\begin{equation}
P G_{\bf r} P = \prod_{\ell \in c} X_\ell \text{,}
\end{equation}
where $c$ is the dual lattice cube centered at ${\bf r}$, and the product is over the 12 edges of the cube.  This is precisely the cube term of the X-cube model.  Moreover,
\begin{equation}
F_{\mu \mu} = \prod_{\ell \sim {\bf s} , \, \ell \perp \mu} Z_\ell \text{,}
\end{equation}
where the product is over the four links touching the dual lattice site ${\bf s}$ that are perpendicular to the $\mu$-direction; this is a vertex term of the X-cube model.  Therefore, the Hamiltonian projected to the degenerate subspace is precisely the X-cube model.

\begin{figure}[h]
\includegraphics[width=.4\textwidth]{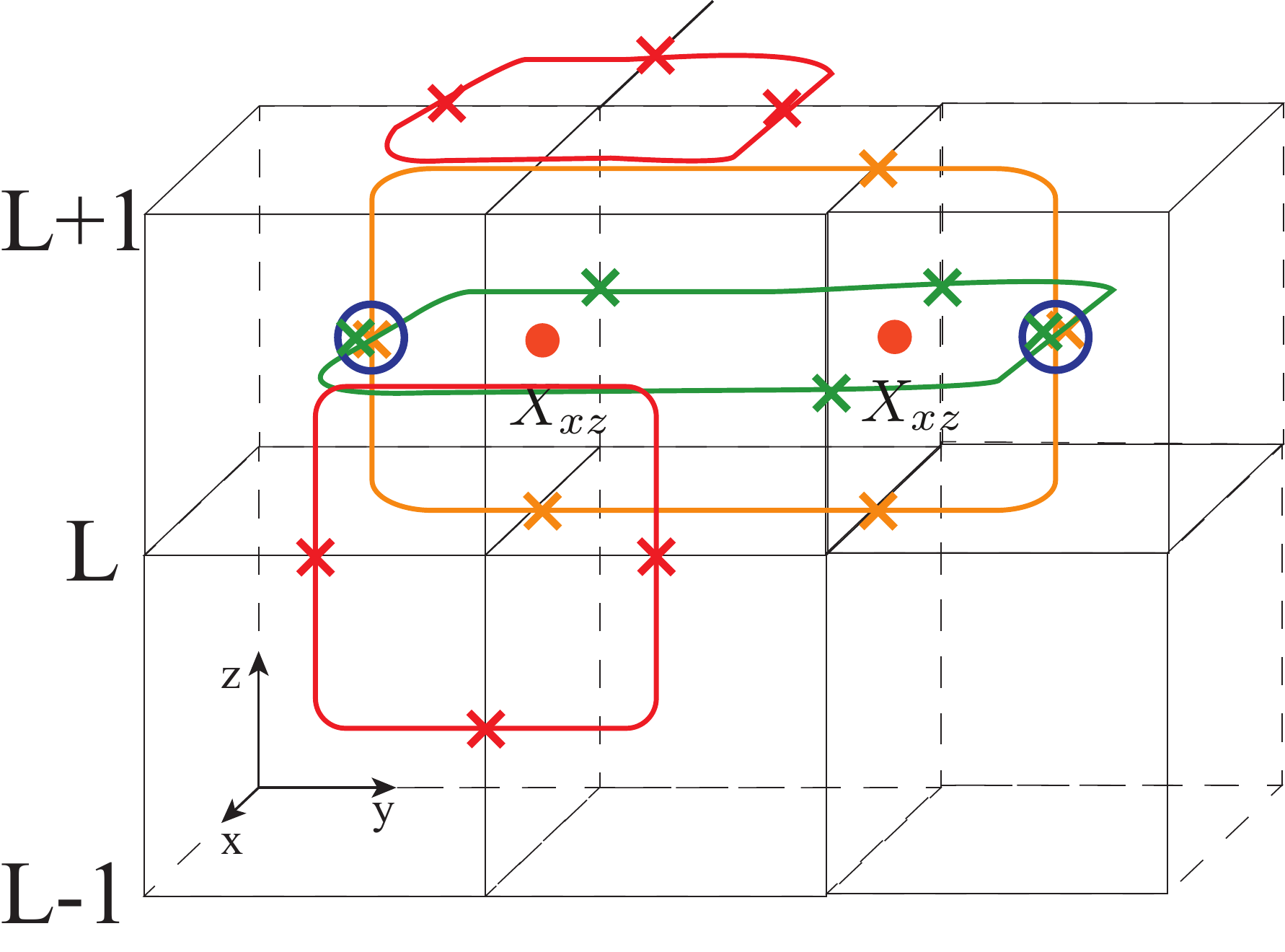}
\caption{The red loops depict the selective flux loop condensate in the rank-2 $\mathbb{Z}_2$ scalar charge with $h$ sufficiently large.  In the presence of the loop condensate, acting with a string of $X_{xz}$ operators over a line extending in the $y$-direction creates two gapped excitations (open circles) at the ends of the string.  Acting with the same operator in the deconfined phase of the rank-2 gauge theory (\emph{i.e.} with $h$ small) also creates excitations along the string depicted by the orange and green loops.  When the loops are condensed, these excitations disappear into the condensate, with only the gapped excitations at the ends of the string remaining. \label{fig:x-cube-transition}}
\end{figure}

\begin{figure}[h]
\includegraphics[width=.4\textwidth]{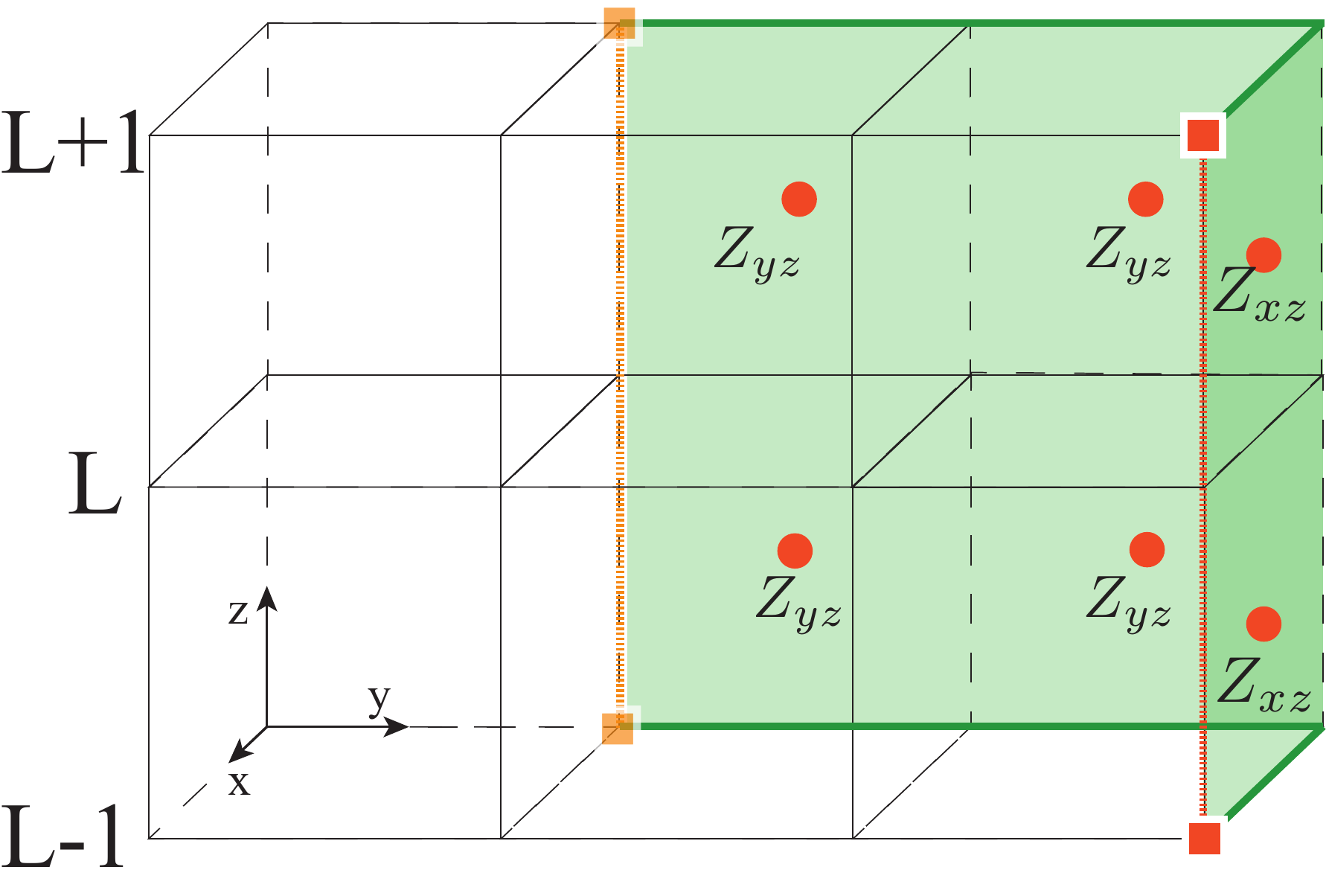}
\caption{The green shaded plaquettes represent an operator transporting a bound state of two charges aligned along the $z$-axis from one location (orange squares) to another (red squares). The operator is a product of $Z_{\mu \nu}$ over the shaded plaquettes. This operator can be viewed as a string operator transporting these bound states within an $xy$-plane, or as a membrane operator creating individual fracton excitations at the corners.
\label{fig:x-cube-pair-fracton}}
\end{figure}

Again, we can develop a physical picture based on flux loop condensation by considering small and intermediate $h$.  Acting with the perturbation creates certain flux loops within $xy$, $yz$ and $xz$ planes, as shown in Fig.~\ref{fig:x-cube-transition}.  When these loops condense, the charge excitations of the rank-2 $\mathbb{Z}_2$ scalar charge theory are confined, as they have a statistical interaction with the loops.  However, bound states of two charges separated along the $x$, $y$ or $z$ axis can propagate freely (see Fig.~\ref{fig:x-cube-pair-fracton}); string operators transporting such bound states are a product of off-diagonal $Z_{\mu \nu}$ operators, which commute with the perturbation.  We thus arrive at the conclusion that individual charges are now created at the corners of a rectangular membrane operator, and become fractons.

We can also obtain the one-dimensional particles of the X-cube model from flux loops in this picture.  Acting on a plaquette $p$ with an off-diagonal operator such as $X_{xz}(p)$ in the rank-2 $\mathbb{Z}_2$ scalar charge theory creates an excitation where eight different $F_{\mu \nu}$ operators (crosses in Fig.~\ref{fig:x-cube-transition}) flip sign.  Four of these are off-diagonal operators with $\mu \neq \nu$, and upon condensing loops these excitations disappear into the condensate.  The other four operators are the $F_{xx}$ and $F_{zz}$ terms adjacent to the plaquette $p$, which remain as gapped excitations.  By acting with a product of $X_{xz}$ over a stack of $xz$-plane plaquettes (see Fig.~\ref{fig:x-cube-transition}), these two excitations can be separated along a line.  These are the one-dimensional particle excitations of the X-cube model.

\section{X-cube model via hollow U(1) scalar charge theory}
\label{sec:hollow}

Now we study the right-hand path of Fig.~\ref{fig:higgs}, which is a different route to the X-cube model from the rank-2 ${\rm U}(1)$ scalar charge theory.  In Sec.~\ref{sec:rank2_U1}, we already noted that adding $U_d \sum_{{\bf r}, \mu} E^2_{\mu \mu}$ to the Hamiltonian of the scalar charge theory, and making $U_d$ large, freezes out the diagonal degrees of freedom, resulting precisely in the hollow ${\rm U}(1)$ gauge theory.  Evidently this is some kind of partial confinement transition, but by what mechanism does it proceed?  Here, in Sec.~\ref{sec:partial-confinement}, we carry out an electric-magnetic duality transformation and show that increasing $U_d$ drives the condensation of certain point-like magnetic monopoles, and that the hollow ${\rm U}(1)$ gauge theory describes the system in the presence of the monopole condensate. The same hollow U(1) scalar charge theory was studied before\cite{xu2008resonating}, and it was found that it does not have a stable deconfined phase.  Nonetheless, the effect of condensing charge-2 matter can still be studied as a mechanism to drive the system into another phase, and in Sec.~\ref{sec:xcube-from-hollowU1} we show that this results in the X-cube model.  Because of the confinement in the hollow ${\rm U}(1)$ gauge theory, we can interpret the condensation described in Sec.~\ref{sec:xcube-from-hollowU1}  as a mechanism for a transition between a confined phase and a phase with the topological order of the X-cube model.

\subsection{Magnetic monopole condensation in the rank-2 U(1) scalar charge theory \label{sec:partial-confinement}}

In order to understand the magnetic sector of the rank-2 U(1) scalar charge theory, we go to its dual theory. We will work with a discrete-time Euclidean action that can be derived from  the lattice Hamiltonian of Eq.~(\ref{eqn:h}).  Since we are interested only in the magnetic sector, we drop the coupling to matter fields and consider a pure gauge theory.  The Euclidean action takes the form
\begin{eqnarray}
S &=& i \sum_{\tau, {\bf r}} \sum_{\mu \leq \nu} E_{\mu  \nu} \Delta_\tau A_{\mu \nu} - i \sum_{\tau, {\bf r}} \phi(\tau,{\bf r}) \sum_{\mu \leq \nu} \Delta_{\mu} \Delta_{\nu} E_{\mu \nu}  \nonumber \\ &+&  \sum_{\tau, {\bf r}} \Big[ u_d \sum_{\mu} E_{\mu \mu}^2 +  u_{od} \sum_{\mu < \nu} E_{\mu \nu}^2 \Big]  
- k \sum_{\tau, {\bf r}} \sum_{\mu, \nu} \cos( B_{\mu \nu} ) \text{.}
\end{eqnarray}
Here, pairs $\tau, {\bf r}$ label space-time points, $E_{\mu \nu}$ is an integer-valued field residing on sites ($\mu = \nu$) and plaquettes ($\mu \neq \nu$) of the spatial lattice, and $A_{\mu \nu}$ is a $2\pi$-periodic variable residing in the same locations.  The magnetic field tensor $B_{\mu \nu}$ was defined in terms of $A_{\mu \nu}$ in Sec.~\ref{sec:rank2_U1}, is not symmetric but is traceless, and resides on sites and plaquettes of the spatial dual cubic lattice.  The field $\phi(\tau,{\bf r})$ is $2\pi$-periodic  Lagrange multiplier that imposes the Gauss' law constraint.

Integrating out $\phi({\bf r}, \tau)$, the Gauss' law constraint is solved by writing $E_{\mu \nu}$ in terms of a dual gauge field $\alpha_{\mu \nu}$ taking values in $2\pi {\mathbb Z}$:
\begin{eqnarray}
E_{\mu \mu} &=& \frac{1}{2 \pi} \epsilon_{\mu \lambda \sigma} \Delta_{\lambda} \alpha_{\sigma \mu}  \text{  (no sum over } \mu \text{)} \label{eqn:Ealpha1} \\
E_{\mu \nu} &=&  \frac{1}{2\pi} \big( \epsilon_{\mu \lambda \sigma} \Delta_{\lambda} \alpha_{\sigma \nu} + \epsilon_{\nu \lambda \sigma}\Delta_{\lambda} \alpha_{\sigma \mu}  \big) \quad ( \mu \neq \nu ) \text{.}\label{eqn:Ealpha2}
\end{eqnarray}
Here, $\alpha_{\mu\nu}$ lives on  dual lattice sites if $\mu=\nu$ and on dual plaquettes if $\mu \neq \nu$.  The field $\alpha_{\mu \nu}$ transforms under dual gauge transformations by
\begin{equation}
\alpha_{\mu \nu} \to \alpha_{\mu \nu} + \Delta_{\mu} \lambda_{\nu} + \delta_{\mu \nu} f \text{,}
\end{equation}
where $\lambda_\mu$ lives on the dual lattice links and $f$ lives on dual sites, and both $\lambda_{\mu}$ and $f$ take values in $2\pi \mathbb{Z}$. 

Next, we proceed to the Villain representation for the cosine terms in the action, \emph{i.e.} we replace
\begin{eqnarray}
-k\cos( B_{\mu \nu} ) \rightarrow \frac{1}{2 k} D_{\mu \nu}^2+ i D_{\mu \nu} B_{\mu \nu} 
\end{eqnarray}
where $D_{\mu \nu}$ is a new integer valued field that is summed over.  Similarly to $\alpha_{\mu \nu}$,  $D_{\mu \nu}$ is a tensor field living on the dual lattice sites and plaquettes.

Now we can integrate out $A_{\mu\nu}$ field, which results in a constraint on $D_{\mu\nu}$, that is solved by
\begin{equation}
D_{\mu \nu} = \frac{1}{2\pi} ( \Delta_\tau \alpha_{\mu \nu} - \Delta_{\mu} \psi_{\nu}  - \delta_{\mu \nu} \theta) \text{.}
\end{equation}
Here,  invariance of $D_{\mu \nu}$ under dual gauge transformations requires introducing $2 \pi \mathbb{Z}$-valued fields $\psi_{\mu}$ and $\theta$ living on the links and sites of the dual lattice, respectively.  These fields transform under dual gauge transformations by
\begin{equation}
\begin{aligned}
\psi_{\mu} &\to \psi_{\mu} + \Delta_\tau \lambda_{\mu} \text{,} \\
\theta &\to \theta + \Delta_\tau f \text{.}
\end{aligned}
\end{equation} 

We thus obtain the dual action
\begin{eqnarray}
S_{{\rm dual}} &=& \frac{1}{4 \pi^2} \sum_{\tau, r} \Big[ u_d \sum_{\mu} (\epsilon_{\mu \lambda \sigma} \Delta_{\lambda} \alpha_{\sigma \mu})^2 \nonumber
\\ &+& u_{od} \sum_{\mu < \nu} (\epsilon_{\mu \lambda \sigma} \Delta_{\lambda} \alpha_{\sigma \nu} + \epsilon_{\nu \lambda \sigma}\Delta_{\lambda} \alpha_{\sigma \mu} )^2 \Big] \nonumber \\
&+& \frac{1}{8 \pi^2 k} \sum_{\tau, r} \sum_{\mu, \nu} (  \Delta_\tau \alpha_{\mu \nu} - \Delta_{\mu} \psi_{\nu} - \delta_{\mu \nu} \theta)^2 \text{.} \label{eq:S-dual}
\end{eqnarray}

To proceed, we promote the discrete fields to real-valued fields and introduce cosine terms to softly restore the discreteness constraint.  It is convenient to allow for real-valued dual gauge transformations; in order to do this and keep the cosine terms gauge invariant, we need to introduce new compact fields  $\phi_\mu$ on  dual  links and  $\gamma$  on  dual  sites.  These fields transform as $\phi_\mu \rightarrow \phi_\mu +\lambda_\mu$ and $\gamma \rightarrow \gamma+f$ under dual gauge transformations. We add the cosine terms
\begin{eqnarray}
S' &=& 
- t_m \sum_{\tau, r} \sum_{\mu, \nu} \cos(\Delta_{\mu} \phi_\nu  - \alpha_{\mu \nu} -\gamma \delta_{\mu\nu}) \nonumber \\
&-& t''_m \sum_{\tau, r} \sum_{\mu} \cos(\Delta_{\tau} \phi_\mu - \psi_{\mu} ) \nonumber \\
&-& t_{\theta} \sum_{\tau, r} \cos(\Delta_\tau \gamma - \theta) 
\text{.} \label{eq:S-dual-soft} 
\end{eqnarray}
The discreteness of $\theta$ will not play an important role in our discussion, so we take $t_\theta = 0$ and integrate out $\gamma$ to obtain the action
\begin{eqnarray}
S'_{\rm dual} &=& S_{\rm dual}
- t_m \sum_{\tau, r} \sum_{\mu \neq \nu} \cos(\Delta_{\mu} \phi_\nu  - \alpha_{\mu \nu} ) \nonumber \\
&-& t'_m \sum_{\tau, r} \sum_{\mu,\nu} \cos (\Delta_{\mu} \phi_\mu  - \Delta_{\nu} \phi_\nu -\alpha_{\mu \mu}+ \alpha_{\nu \nu}) \nonumber \\
&-& t''_m \sum_{\tau, r} \sum_{\mu} \cos(\Delta_{\tau} \phi_\mu - \psi_{\mu} ) \text{.} \label{eq:S-dual-soft} 
\end{eqnarray}
Here, we see that $\phi_{\mu}$ is a matter field coupled to the dual gauge field.  We therefore expect it to represent gapped magnetic excitations, as we shall see below.

Next, we decouple the second term in Eq.~(\ref{eq:S-dual}) with a Hubbard-Stratonovich transformation,
\begin{eqnarray}
\frac{1}{8 \pi^2 k}  &(&\Delta_\tau \alpha_{\mu \nu} - \Delta_{\mu} \psi_{\nu} - \delta_{\mu \nu} \theta)^2 \nonumber \\   &\rightarrow& 2 \pi^2 k   \beta_{\mu \nu}^2 - i \beta_{\mu \nu} ( \Delta_\tau \alpha_{\mu \nu} - \Delta_{\mu} \psi_{\nu} - \delta_{\mu \nu} \theta)   \text{.}
\end{eqnarray}
Here, $\beta_{\mu \nu}$ should be interpreted as the magnetic field; that is, $\beta_{\mu \nu} \sim B_{\mu \nu}$.  We note that integrating out $\theta$ gives the constraint $\sum_{\mu} \beta_{\mu \mu} = 0$.
Next, we go to a Villain representation for the last term in Eq.~(\ref{eq:S-dual-soft}), \emph{i.e.} 
\begin{eqnarray}
t''_m \cos&(&\Delta_{\tau} \phi_\mu - \psi_{\mu} ) \rightarrow  \frac{1}{t''_m}  n_{\mu}^2
+ i  n_\mu ( \Delta_\tau \phi_{\mu} - \psi_{\mu} ) 
\end{eqnarray}
This introduces the ``number operator'' $n_{\mu}$ conjugate to the phase $\phi_{\mu}$ of the dual matter field.

Integrating out $\psi_{\mu}$ gives the dual Gauss' law,
\begin{equation}
\Delta_\mu \beta_{\mu\nu} = n_\nu \text{,}
\end{equation}
which shows that $n_\mu$ is a point-like (in space) vector magnetic charge, or magnetic monopole, created by $e^{i \phi_{\mu}}$.  From this constraint, it is straightforward to show that the quantity $\sum_{\mu} r_{\mu} n_{\mu}$ is conserved.  Conservation of this quantity allows a $n_{\mu}$ monopole to hop freely in a $d=2$ plane normal to $\mu$, but forbids the monopole to move independently in the $\mu$-direction.  Therefore, these monopoles are two-dimensional particles.  This is born out by inspecting the the first two terms in Eq.~(\ref{eq:S-dual-soft}), which describe two different dynamical processes of monopoles.  The $t_m$ term is a hopping of $n_{\mu}$ monopoles in a plane perpendicular to $\mu$.  The $t'_m$ term is a cooperative two-monopole hopping process.

We now consider the effect of condensing monopoles within their $d=2$ planes of motion, by making $t_m$ and $t''_m$ large, while keeping $t'_m$ fixed.  We will see that, because we are keeping $t'_m$ fixed, this does not fully Higgs out the dual gauge field.  To understand why this is the partial confinement transition we expect when $U_d$ (and hence $u_d$) becomes large, we note that the $t_m$ cosine originates from the discreteness constraint on off-diagonal elements of $\alpha_{\mu \nu}$.  Now, diagonal elements of $E_{\mu \nu}$ only depend on off-diagonal elements of $a_{\mu \nu}$, in contrast to off-diagonal elements of $E_{\mu \nu}$, which depend on both diagonal and off-diagonal elements of $a_{\mu \nu}$ [see Eqs.~\ref{eqn:Ealpha1} and~\ref{eqn:Ealpha2}].  Therefore, as $U_d$ is increased in the original Hamiltonian, the discreteness of diagonal elements $E_{\mu \mu}$ becomes more important, and we expect $t_m$ to increase in this effective dual description.  

To analyze the large $t_m$ and $t''_m$ limit, we make the changes of variables
\begin{eqnarray}
\alpha_{\mu \nu} &\to& \alpha_{\mu \nu} + \Delta_{\mu} \phi_{\nu}  \\
\psi_{\mu} &\to& \psi_{\mu} + \Delta_{\tau} \phi_{\mu} \text{.}
\end{eqnarray}
We thus obtain constraints $\alpha_{\mu \nu} = 0$ (for $\mu \neq \nu$), and $\psi_{\mu} = 0$.  The first of these constraints implies $E_{\mu \mu} = 0$, as expected in the large $U_d$ limit.   We define $h_\mu \equiv \alpha_{\mu\mu}$, and write the resulting action as
\begin{eqnarray}
S^{{\rm hollow}}_{{\rm dual}} &=& \frac{u_{od}}{4 \pi^2} \sum_{\tau, r} \Big[ (\Delta_z h_x - \Delta_z h_y)^2
+ (\Delta_x h_y - \Delta_x h_z)^2 \nonumber \\ &+& (\Delta_y h_z - \Delta_y h_x)^2 \Big]
+ \frac{1}{8 \pi^2 k} \sum_{\tau, r} \sum_{\mu} ( \Delta_\tau h_{\mu} - \theta )^2  \nonumber  \\
&-& t'_m \sum_{\tau, r} \sum_{\mu < \nu} \cos( h_{\mu} - h_{\nu}  )  \text{.}
\end{eqnarray}
which is invariant under the gauge transformations
\begin{eqnarray}
h_{\mu} &\to& h_{\mu} + f \\
\theta &\to& \theta + \Delta_\tau f \text{,}
\end{eqnarray}

It was shown previously by Xu and Wu that this theory is a dual description of the hollow ${\rm U}(1)$ scalar charge theory,\cite{xu2008resonating} so that this theory indeed describes the system upon condensing monopoles within their $d=2$ planes of motion.  Moreover, Xu and Wu also showed that the cosine term in the above action is relevant, corresponding to a proliferation of instantons in space-time, that results in a gapped confined phase.\cite{xu2008resonating}

\subsection{X-cube model from lattice hollow ${\rm U}(1)$ scalar charge theory by Higgs mechanism}
\label{sec:xcube-from-hollowU1}

Here, we couple the hollow ${\rm U}(1)$ rank-2 scalar charge theory to charge-2 matter, and show that condensing this matter results in the X-cube model.  The details of the analysis are very similar to that carried out in Sec.~\ref{sec:charge-2}, so we only summarize the key points.  Starting with the Hamiltonian Eq.~(\ref{eqn:h_hollow}), we add a charge-2 matter field with number $N_{\bf r}$ and conjugate phase $\Theta_{\bf r}$.  We add the following terms to the Hamiltonian
\begin{equation}
\begin{aligned}
H_{2e} &= u_2 \sum_{\bf r} N_{\bf r}^2 - \Delta \sum_{{\bf r}} \cos \left[ \Theta_{\bf r} -2 \theta_{\bf r} \right] \\
& - J_2 \sum_{{\bf r},\mu < \nu} \cos \biggl[ \Delta_{\mu} \Delta_{\nu} \Theta
-2 A_{\mu\nu}({\bf r})\biggr] \text{.}
\end{aligned}
\end{equation}
The charge-2 matter condenses for sufficiently large $J_2$.  As in Sec.~\ref{sec:charge-2}, we take both $J_2$ and $\Delta$ to be large, which allows us to treat the corresponding cosine terms as constraints.

We obtain a model defined in terms of symmetric-tensor $Z_{\mu \nu}, X_{\mu \nu}$ Pauli operators with $\mu \neq \nu$ living on plaquettes, and $\tau^x_{\bf r}, \tau^z_{\bf r}$ Pauli operators living on sites.  As in Sec.~\ref{sec:charge-2}, $X_{\mu \nu}$ is a rank-2 $\mathbb{Z}_2$ electric field, $Z_{\mu \nu}$ the conjugate rank-2 $\mathbb{Z}_2$ gauge field, and $\tau^x_{\bf r}$ is the $\mathbb{Z}_2$ gauge charge of the un-condensed charge-1 matter.  The Hamiltonian is
\begin{eqnarray}
H^{{\rm hollow}}_{\mathbb{Z}_2} &=& - J \sum_{p, \mu < \nu} \tau^z_{{\bf r}_1} \tau^z_{{\bf r}_2} \tau^z_{{\bf r}_3} \tau^z_{{\bf r}_4}  Z_{\mu \nu}(p) \nonumber \\
&-& K \sum_{{\bf r}, \mu} F_{\mu \mu} - u \sum_{\bf r} \tau^x_{\bf r} - U \sum_{{\bf r}, \mu < \nu} X_{\mu \nu} \text{.}
\end{eqnarray}
Here, the sum in the first term is over plaquettes $p$, and the sites ${\bf r}_1, \dots, {\bf r}_4$ lie at the corners of $p$.  $F_{\mu \nu}$ is defined in Sec.~\ref{sec:charge-2}; only the diagonal elements play a role here, because only the diagonal elements $B_{\mu \mu}$ appear in the ${\rm U}(1)$ hollow gauge theory.  The Gauss' law is
\begin{equation}
G^h_{\bf r} \equiv \prod_{p \sim {\bf r}} X_{\mu \nu}(p) =  \tau^x_{\bf r} \text{,}
\end{equation}
where the product is over the 12 plaquettes sharing a corner with ${\bf r}$.

\begin{figure}[h]
\includegraphics[width=.45\textwidth]{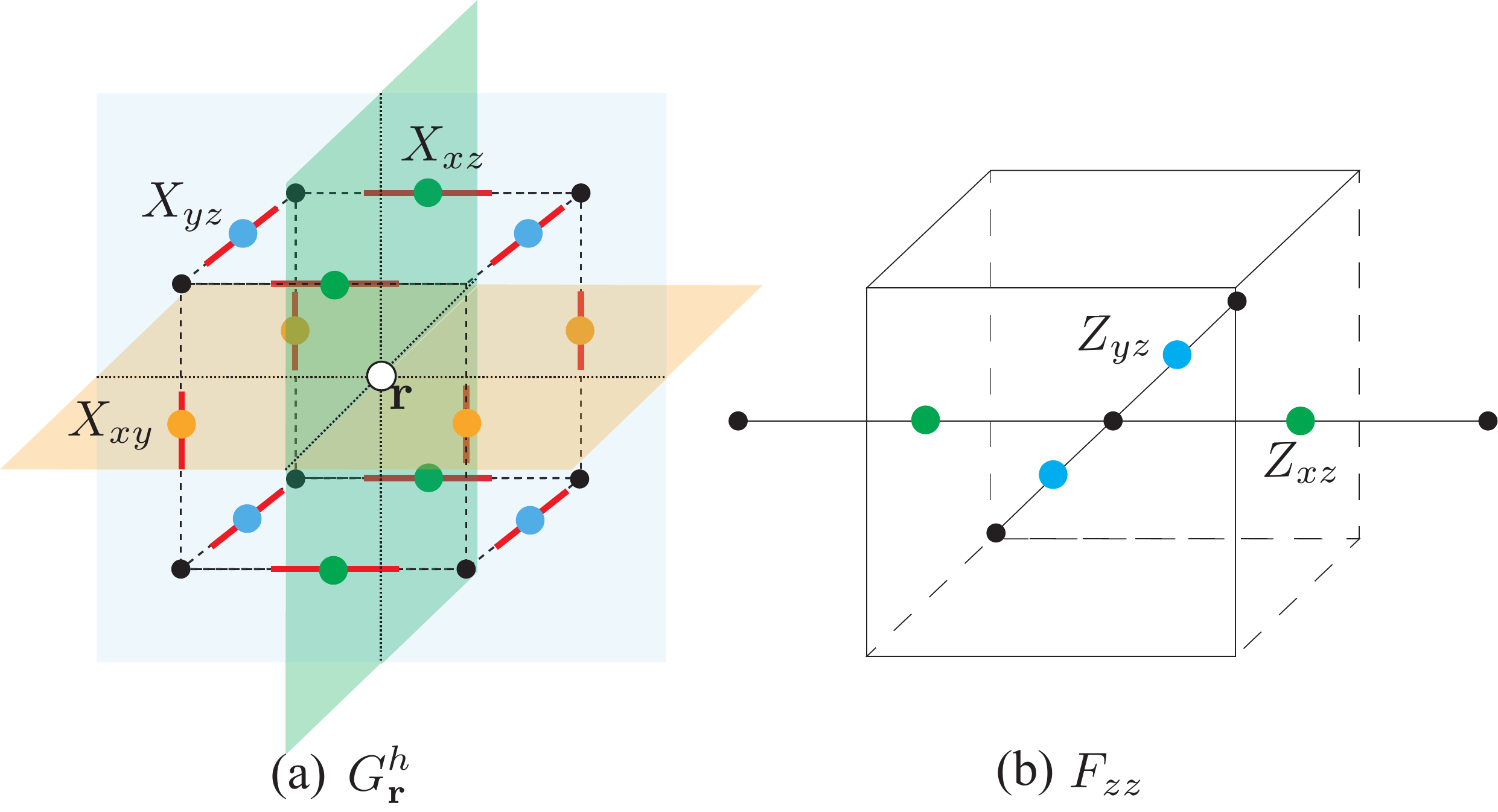}
\caption{(a) The operator $G^h_{\bf r}$ is a product of $X_{\mu \nu}$ over  the plaquettes touching ${\bf r}$ (shaded plaquettes).  Viewing $X_{\mu \nu}$ as a link variable on the dual lattice, $G^h_{\bf r}$ is a product over the edges of the dual lattice cube surrounding ${\bf r}$ (dashed lines).  (b)  Diagonal components of $F_{\mu \nu}$, such as $F_{zz}$ illustrated here, are products of $Z_{\mu \nu}$ over four of the faces surrounding a cube.  On the dual lattice, this becomes a product over  four links touching a dual site, as shown.  \label{fig:X-cube_dual}}
\end{figure}

This theory can be viewed as a hollow rank-2 ${\mathbb Z}_2$ scalar charge theory on the cubic lattice.  To see it is equivalent to the X-cube model, we exploit the same mapping to a local bosonic model discussed in Sec.~\ref{sec:charge-2}.  The resulting Hamiltonian is 
\begin{eqnarray}
\tilde{H}^{{\rm hollow}}_{\mathbb{Z}_2} &=& - J \sum_{p, \mu < \nu}  \tilde{Z}_{\mu \nu}(p) - K \sum_{{\bf r}, \mu} \tilde{F}_{\mu \mu} \nonumber \\
&-& u \sum_{\bf r} \tilde{G}^h_{\bf r} - U \sum_{{\bf r}, \mu < \nu} \tilde{X}_{\mu \nu} \text{.}
\end{eqnarray}
Passing to the dual lattice so that the plaquette variables become link variables, the $\tilde{F}_{\mu \mu}$ term becomes precisely the vertex term of the X-cube model, while the $\tilde{G}^h$ term becomes the cube term (see Fig.~\ref{fig:X-cube_dual}).  Therefore, we have obtained precisely the X-cube model, perturbed by the $J$ and $U$ terms, which give dynamics to the fractons and one-dimensional particle excitations, respectively.

\section{Checkerboard model as a $\mathbb{Z}_2$ scalar charge model \label{sec:checkerboard}}

All the rank-2 gauge theories considered thus far have a Gauss' law of the same form at every point in space.  Here, we construct a rank-2 $\mathbb{Z}_2$ gauge theory with two different Gauss' laws at different vertices of the cubic lattice, and show that this theory is equivalent to the checkerboard fracton model.\cite{vijay2016fracton}.

We work on the cubic lattice, and place qubits on $xz$ and $yz$ plaquettes, with no degrees of freedom residing on $xy$ plaquettes.  We denote the corresponding Pauli operators by $X_{xz (yz)}$ and $Z_{xz (yz)}$.  It is convenient to work with electric field $E_{\mu \nu}$ and gauge field $A_{\mu \nu}$ variables introduced by defining $X_{\mu \nu} = \exp(i \pi E_{\mu \nu} )$, and $Z_{\mu \nu} = \exp(i \pi A_{\mu \nu})$, where $E_{\mu \nu}$ and $A_{\mu \nu}$ take values in $\{ 0, 1\}$.  We write
\begin{equation}
E_{\mu \nu} = \left[ \begin{array}{ccc}
0 & E_{xz} & E_{yz} \\
E_{zx} & 0 & 0 \\
E_{yz} & 0 & 0
\end{array} \right] \quad \quad A_{\mu \nu} = \left[ \begin{array}{ccc}
0 & A_{xz} & A_{yz} \\
A_{zx} & 0 & 0 \\
A_{yz} & 0 & 0
\end{array} \right] 
\end{equation}
Taken together, the centers of the $xz$ and $yz$ plaquettes form a finer simple cubic lattice.

\begin{figure}[h]
\includegraphics[width=.5\textwidth]{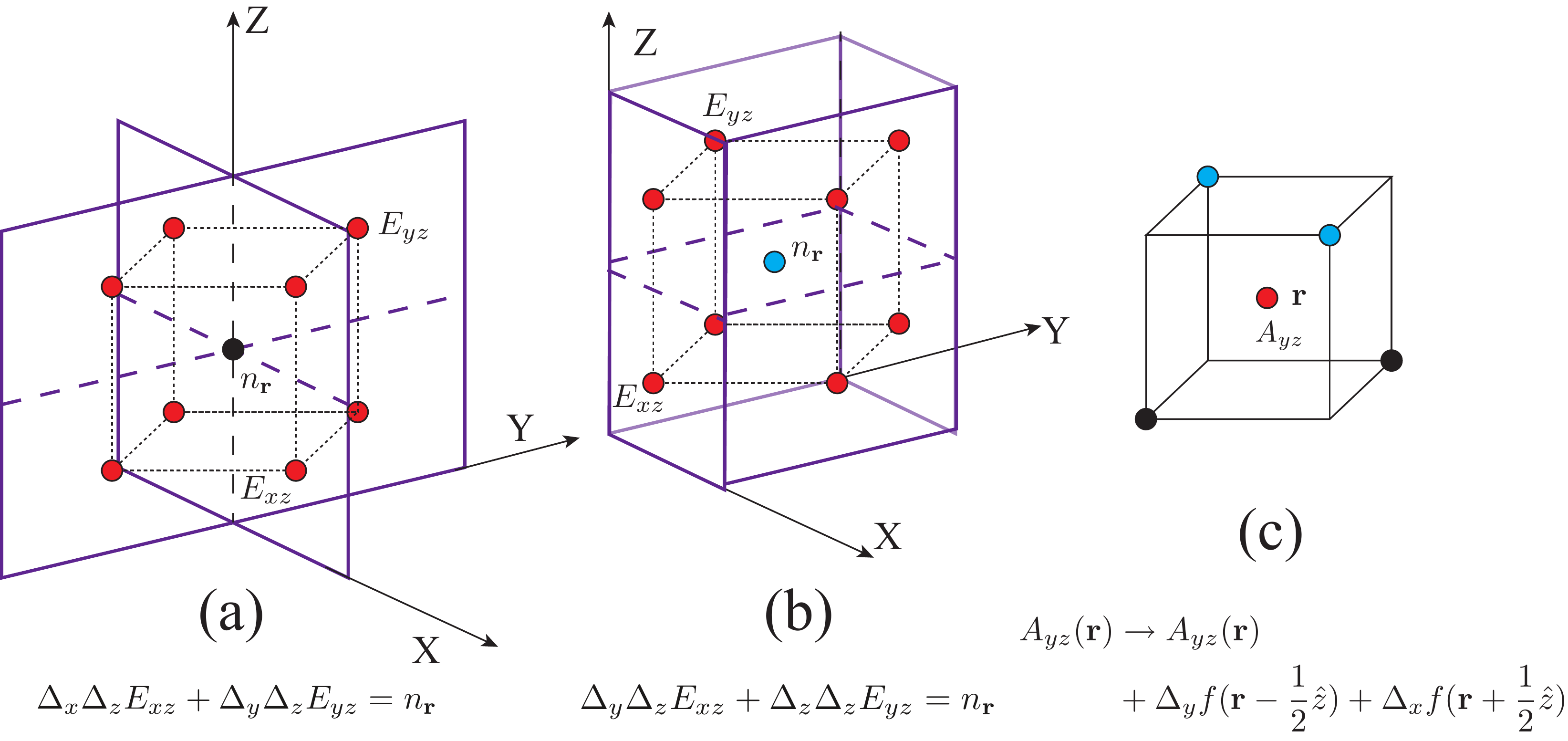}
\caption{(a) The Gauss law for the even layers involves eight electric field variables forming a cube surrounding an even-$z$ site of the cubic lattice (black circle).  (b) The Gauss law for the odd layers is a sum of the electric field over the eight vertices of a cube surrounding the center of an $xy$-plane plaquette whose center has odd $z$ coordinate.  (c) Gauge transformation of $A_{yz}$.  The black and blue circles show the locations of the nearby checkerboard cube centers where the function $f$ is defined. \label{fig:ckb}}
\end{figure}

We impose two  Gauss' laws.  On layers with even $z$ coordinate, we impose $\Delta_x \Delta_z E_{xz} + \Delta_y \Delta_z E_{yz} = n_{\bf r}$.  The corresponding charge resides at even-$z$ vertices of the cubic lattice, as shown in Fig.~\ref{fig:ckb}a.  On odd-$z$ layers, we impose $\Delta_y \Delta_z E_{xz} + \Delta_x \Delta_z E_{yz} = n_{\bf r'}$, where $\bf r'$ is the center of an $xy$ plaquette, so that charge resides on $xy$ plaquettes (see Fig.~\ref{fig:ckb}b).  Viewing these Gauss' laws in terms of the finer cubic lattice, we see that they are defined on a ``checkerboard'' of edge-sharing cubes whose centers (where charges reside) form a FCC lattice.

From the Gauss law, we can obtain the gauge transformation of $A_{\mu \nu}$, \emph{e.g.} $A_{yz} ({\bf r}) \rightarrow A_{yz}({\bf r}) + \Delta_y f({\bf r}- \frac{1}{2} \hat{z}) + \Delta_x f({\bf r}+ \frac{1}{2} \hat{z})$ as shown in Fig.~\ref{fig:ckb}(d). The simplest gauge-invariant combinations of the $A_{\mu \nu}$ are sums of eight variables that take precisely the same form as the sums in the two Gauss' laws.  Passing to a local bosonic model as above, the Hamiltonian consists of a term imposing the  Gauss' laws, and the gauge-invariant terms built from $A_{\mu \nu}$.  In terms of Pauli matrices, 
we have
\begin{equation}
H_{ckb} = -\sum_C\prod_{i \in C } X_i -\sum_C\prod_{i \in C } Z_i
\end{equation}
where $C$ denotes the edge-sharing cubes of the checkerboard lattice.  This is precisely the Hamiltonian of the checkerboard fracton model.

\section{Discussion}
\label{sec:discussion}

In this work, we showed that Higgs and partial confinement mechanisms provide relationships among gapped fracton phases and ${\rm U}(1)$ symmetric-tensor gauge theories.  These relationships also encompass more conventional topological orders, \emph{e.g.} the rank-2 $\mathbb{Z}_2$ scalar charge theory that is obtained from the rank-2 ${\rm U}(1)$ scalar charge theory by condensing charge-2 matter, and which becomes the X-cube fracton topological order upon selective loop condensation. Our results are a starting point to investigate quantum critical phenomena at transitions between different fracton states.  At a more basic level, the mechanisms we discuss give insight into fracton phases, by allowing us to understand the degrees of freedom in terms of another proximate phase.

In this work, we focused on the scalar charge theory, but there are other rank-2 U(1) gauge theories\cite{pretko2016subdimensional} with different forms of Gauss' law, \emph{e.g.} the vector charge theory and traceless scalar charge theories. It will be interesting to investigate what types of fracton topological order can arise from more general higher-rank gauge theories, via Higgs mechanisms and partial confinement transitions. In addition, it may be fruitful to explore whether new types of fracton topological order can arise as gauge theories with different forms of Gauss' law on different lattice sites, along the lines of our presentation of the checkerboard fracton model as a rank-2 gauge theory.

Another open question is whether any ``type II'' fracton phases like that in Haah's code, where all non-trivial excitations are fractons created at the corners of fractal operators, are  related to ${\rm U}(1)$ symmetric-tensor gauge theory in a manner similar to the examples discussed here.  In a certain sense, it is obvious that Haah's code can be viewed as a gauge theory, where one type of term in the Hamiltonian (either the $X$ or $Z$ term) is viewed as implementing a Gauss' law constraint energetically.  The question then becomes whether this Gauss' law can emerge from simpler or more familiar theories by some sequence of Higgs and/or partial confinement transitions, which could shed light on type II fracton topological order beyond the realm of exactly solvable models.

\section*{Acknowledgement}

H.M thanks Michael Pretko for insightful discussion. M.H. and H.M. are supported by the U.S. Department of Energy, Office of Science, Basic Energy Sciences (BES) under Award number DE-SC0014415. X.C. is supported by the Caltech Institute for Quantum Information and Matter, the Walter Burke Institute for Theoretical Physics, the Alfred P. Sloan research fellowship, and National Science Foundation under award number DMR-1654340.

{\bf Note added:} We became aware of a parallel investigation \cite{bulmash2018higgs}
in which the Higgs mechanism is applied to several rank-2 gauge theories and the resulting phases are studied. Where our results overlap, they agree.

\bibliography{ref}

\begin{thebibliography}{40}%
\makeatletter
\providecommand \@ifxundefined [1]{%
 \@ifx{#1\undefined}
}%
\providecommand \@ifnum [1]{%
 \ifnum #1\expandafter \@firstoftwo
 \else \expandafter \@secondoftwo
 \fi
}%
\providecommand \@ifx [1]{%
 \ifx #1\expandafter \@firstoftwo
 \else \expandafter \@secondoftwo
 \fi
}%
\providecommand \natexlab [1]{#1}%
\providecommand \enquote  [1]{``#1''}%
\providecommand \bibnamefont  [1]{#1}%
\providecommand \bibfnamefont [1]{#1}%
\providecommand \citenamefont [1]{#1}%
\providecommand \href@noop [0]{\@secondoftwo}%
\providecommand \href [0]{\begingroup \@sanitize@url \@href}%
\providecommand \@href[1]{\@@startlink{#1}\@@href}%
\providecommand \@@href[1]{\endgroup#1\@@endlink}%
\providecommand \@sanitize@url [0]{\catcode `\\12\catcode `\$12\catcode
  `\&12\catcode `\#12\catcode `\^12\catcode `\_12\catcode `\%12\relax}%
\providecommand \@@startlink[1]{}%
\providecommand \@@endlink[0]{}%
\providecommand \url  [0]{\begingroup\@sanitize@url \@url }%
\providecommand \@url [1]{\endgroup\@href {#1}{\urlprefix }}%
\providecommand \urlprefix  [0]{URL }%
\providecommand \Eprint [0]{\href }%
\providecommand \doibase [0]{http://dx.doi.org/}%
\providecommand \selectlanguage [0]{\@gobble}%
\providecommand \bibinfo  [0]{\@secondoftwo}%
\providecommand \bibfield  [0]{\@secondoftwo}%
\providecommand \translation [1]{[#1]}%
\providecommand \BibitemOpen [0]{}%
\providecommand \bibitemStop [0]{}%
\providecommand \bibitemNoStop [0]{.\EOS\space}%
\providecommand \EOS [0]{\spacefactor3000\relax}%
\providecommand \BibitemShut  [1]{\csname bibitem#1\endcsname}%
\let\auto@bib@innerbib\@empty
\bibitem [{\citenamefont {Chamon}(2005)}]{chamon2005quantum}%
  \BibitemOpen
  \bibfield  {author} {\bibinfo {author} {\bibfnamefont {Claudio}\ \bibnamefont
  {Chamon}},\ }\bibfield  {title} {\enquote {\bibinfo {title} {Quantum
  glassiness in strongly correlated clean systems: an example of topological
  overprotection},}\ }\href@noop {} {\bibfield  {journal} {\bibinfo  {journal}
  {Physical review letters}\ }\textbf {\bibinfo {volume} {94}},\ \bibinfo
  {pages} {040402} (\bibinfo {year} {2005})}\BibitemShut {NoStop}%
\bibitem [{\citenamefont {Bravyi}\ \emph {et~al.}(2011)\citenamefont {Bravyi},
  \citenamefont {Leemhuis},\ and\ \citenamefont
  {Terhal}}]{bravyi2011topological}%
  \BibitemOpen
  \bibfield  {author} {\bibinfo {author} {\bibfnamefont {Sergey}\ \bibnamefont
  {Bravyi}}, \bibinfo {author} {\bibfnamefont {Bernhard}\ \bibnamefont
  {Leemhuis}}, \ and\ \bibinfo {author} {\bibfnamefont {Barbara~M}\
  \bibnamefont {Terhal}},\ }\bibfield  {title} {\enquote {\bibinfo {title}
  {Topological order in an exactly solvable 3d spin model},}\ }\href@noop {}
  {\bibfield  {journal} {\bibinfo  {journal} {Annals of Physics}\ }\textbf
  {\bibinfo {volume} {326}},\ \bibinfo {pages} {839--866} (\bibinfo {year}
  {2011})}\BibitemShut {NoStop}%
\bibitem [{\citenamefont {Haah}(2011)}]{haah2011local}%
  \BibitemOpen
  \bibfield  {author} {\bibinfo {author} {\bibfnamefont {Jeongwan}\
  \bibnamefont {Haah}},\ }\bibfield  {title} {\enquote {\bibinfo {title} {Local
  stabilizer codes in three dimensions without string logical operators},}\
  }\href@noop {} {\bibfield  {journal} {\bibinfo  {journal} {Physical Review
  A}\ }\textbf {\bibinfo {volume} {83}},\ \bibinfo {pages} {042330} (\bibinfo
  {year} {2011})}\BibitemShut {NoStop}%
\bibitem [{\citenamefont {Bravyi}\ and\ \citenamefont
  {Haah}(2013)}]{bravyi2013quantum}%
  \BibitemOpen
  \bibfield  {author} {\bibinfo {author} {\bibfnamefont {Sergey}\ \bibnamefont
  {Bravyi}}\ and\ \bibinfo {author} {\bibfnamefont {Jeongwan}\ \bibnamefont
  {Haah}},\ }\bibfield  {title} {\enquote {\bibinfo {title} {Quantum
  self-correction in the 3d cubic code model},}\ }\href {\doibase
  10.1103/PhysRevLett.111.200501} {\bibfield  {journal} {\bibinfo  {journal}
  {Phys. Rev. Lett.}\ }\textbf {\bibinfo {volume} {111}},\ \bibinfo {pages}
  {200501} (\bibinfo {year} {2013})}\BibitemShut {NoStop}%
\bibitem [{\citenamefont {Yoshida}(2013)}]{yoshida2013exotic}%
  \BibitemOpen
  \bibfield  {author} {\bibinfo {author} {\bibfnamefont {Beni}\ \bibnamefont
  {Yoshida}},\ }\bibfield  {title} {\enquote {\bibinfo {title} {Exotic
  topological order in fractal spin liquids},}\ }\href@noop {} {\bibfield
  {journal} {\bibinfo  {journal} {Physical Review B}\ }\textbf {\bibinfo
  {volume} {88}},\ \bibinfo {pages} {125122} (\bibinfo {year}
  {2013})}\BibitemShut {NoStop}%
\bibitem [{\citenamefont {Hsieh}\ and\ \citenamefont
  {Hal\'asz}(2017)}]{hsieh2017fractons}%
  \BibitemOpen
  \bibfield  {author} {\bibinfo {author} {\bibfnamefont {Timothy~H.}\
  \bibnamefont {Hsieh}}\ and\ \bibinfo {author} {\bibfnamefont {G\'abor~B.}\
  \bibnamefont {Hal\'asz}},\ }\bibfield  {title} {\enquote {\bibinfo {title}
  {Fractons from partons},}\ }\href {\doibase 10.1103/PhysRevB.96.165105}
  {\bibfield  {journal} {\bibinfo  {journal} {Phys. Rev. B}\ }\textbf {\bibinfo
  {volume} {96}},\ \bibinfo {pages} {165105} (\bibinfo {year}
  {2017})}\BibitemShut {NoStop}%
\bibitem [{\citenamefont {Vijay}\ \emph {et~al.}(2015)\citenamefont {Vijay},
  \citenamefont {Haah},\ and\ \citenamefont {Fu}}]{vijay2015new}%
  \BibitemOpen
  \bibfield  {author} {\bibinfo {author} {\bibfnamefont {Sagar}\ \bibnamefont
  {Vijay}}, \bibinfo {author} {\bibfnamefont {Jeongwan}\ \bibnamefont {Haah}},
  \ and\ \bibinfo {author} {\bibfnamefont {Liang}\ \bibnamefont {Fu}},\
  }\bibfield  {title} {\enquote {\bibinfo {title} {A new kind of topological
  quantum order: A dimensional hierarchy of quasiparticles built from
  stationary excitations},}\ }\href@noop {} {\bibfield  {journal} {\bibinfo
  {journal} {Physical Review B}\ }\textbf {\bibinfo {volume} {92}},\ \bibinfo
  {pages} {235136} (\bibinfo {year} {2015})}\BibitemShut {NoStop}%
\bibitem [{\citenamefont {Vijay}\ \emph {et~al.}(2016)\citenamefont {Vijay},
  \citenamefont {Haah},\ and\ \citenamefont {Fu}}]{vijay2016fracton}%
  \BibitemOpen
  \bibfield  {author} {\bibinfo {author} {\bibfnamefont {Sagar}\ \bibnamefont
  {Vijay}}, \bibinfo {author} {\bibfnamefont {Jeongwan}\ \bibnamefont {Haah}},
  \ and\ \bibinfo {author} {\bibfnamefont {Liang}\ \bibnamefont {Fu}},\
  }\bibfield  {title} {\enquote {\bibinfo {title} {Fracton topological order,
  generalized lattice gauge theory and duality},}\ }\href@noop {} {\bibfield
  {journal} {\bibinfo  {journal} {arXiv preprint arXiv:1603.04442}\ } (\bibinfo
  {year} {2016})}\BibitemShut {NoStop}%
\bibitem [{\citenamefont {Petrova}\ and\ \citenamefont
  {Regnault}(2017)}]{petrova2017simple}%
  \BibitemOpen
  \bibfield  {author} {\bibinfo {author} {\bibfnamefont {Olga}\ \bibnamefont
  {Petrova}}\ and\ \bibinfo {author} {\bibfnamefont {Nicolas}\ \bibnamefont
  {Regnault}},\ }\bibfield  {title} {\enquote {\bibinfo {title} {A simple
  anisotropic three-dimensional quantum spin liquid with fracton topological
  order},}\ }\href@noop {} {\bibfield  {journal} {\bibinfo  {journal} {arXiv
  preprint arXiv:1709.10094}\ } (\bibinfo {year} {2017})}\BibitemShut {NoStop}%
\bibitem [{\citenamefont {Slagle}\ and\ \citenamefont
  {Kim}(2017{\natexlab{a}})}]{slagle2017fracton}%
  \BibitemOpen
  \bibfield  {author} {\bibinfo {author} {\bibfnamefont {Kevin}\ \bibnamefont
  {Slagle}}\ and\ \bibinfo {author} {\bibfnamefont {Yong~Baek}\ \bibnamefont
  {Kim}},\ }\bibfield  {title} {\enquote {\bibinfo {title} {Fracton topological
  order from nearest-neighbor two-spin interactions and continuous
  subdimensional quantum phase transitions via dualities},}\ }\href@noop {}
  {\bibfield  {journal} {\bibinfo  {journal} {arXiv preprint arXiv:1704.03870}\
  } (\bibinfo {year} {2017}{\natexlab{a}})}\BibitemShut {NoStop}%
\bibitem [{\citenamefont {Prem}\ \emph {et~al.}(2017)\citenamefont {Prem},
  \citenamefont {Haah},\ and\ \citenamefont {Nandkishore}}]{prem2017glassy}%
  \BibitemOpen
  \bibfield  {author} {\bibinfo {author} {\bibfnamefont {Abhinav}\ \bibnamefont
  {Prem}}, \bibinfo {author} {\bibfnamefont {Jeongwan}\ \bibnamefont {Haah}}, \
  and\ \bibinfo {author} {\bibfnamefont {Rahul}\ \bibnamefont {Nandkishore}},\
  }\bibfield  {title} {\enquote {\bibinfo {title} {Glassy quantum dynamics in
  translation invariant fracton models},}\ }\href@noop {} {\bibfield  {journal}
  {\bibinfo  {journal} {Physical Review B}\ }\textbf {\bibinfo {volume} {95}},\
  \bibinfo {pages} {155133} (\bibinfo {year} {2017})}\BibitemShut {NoStop}%
\bibitem [{\citenamefont {Slagle}\ and\ \citenamefont
  {Kim}(2017{\natexlab{b}})}]{slagle2017quantum}%
  \BibitemOpen
  \bibfield  {author} {\bibinfo {author} {\bibfnamefont {Kevin}\ \bibnamefont
  {Slagle}}\ and\ \bibinfo {author} {\bibfnamefont {Yong~Baek}\ \bibnamefont
  {Kim}},\ }\bibfield  {title} {\enquote {\bibinfo {title} {Quantum field
  theory of fracton topological order and" topological" degeneracy from
  geometry},}\ }\href@noop {} {\bibfield  {journal} {\bibinfo  {journal} {arXiv
  preprint arXiv:1708.04619}\ } (\bibinfo {year}
  {2017}{\natexlab{b}})}\BibitemShut {NoStop}%
\bibitem [{\citenamefont {Slagle}\ and\ \citenamefont
  {Kim}(2017{\natexlab{c}})}]{slagle2017x}%
  \BibitemOpen
  \bibfield  {author} {\bibinfo {author} {\bibfnamefont {Kevin}\ \bibnamefont
  {Slagle}}\ and\ \bibinfo {author} {\bibfnamefont {Yong~Baek}\ \bibnamefont
  {Kim}},\ }\bibfield  {title} {\enquote {\bibinfo {title} {X-cube model on
  generic lattices: New phases and geometric order},}\ }\href@noop {}
  {\bibfield  {journal} {\bibinfo  {journal} {arXiv preprint arXiv:1712.04511}\
  } (\bibinfo {year} {2017}{\natexlab{c}})}\BibitemShut {NoStop}%
\bibitem [{\citenamefont {Shirley}\ \emph {et~al.}(2017)\citenamefont
  {Shirley}, \citenamefont {Slagle}, \citenamefont {Wang},\ and\ \citenamefont
  {Chen}}]{shirley2017fracton}%
  \BibitemOpen
  \bibfield  {author} {\bibinfo {author} {\bibfnamefont {Wilbur}\ \bibnamefont
  {Shirley}}, \bibinfo {author} {\bibfnamefont {Kevin}\ \bibnamefont {Slagle}},
  \bibinfo {author} {\bibfnamefont {Zhenghan}\ \bibnamefont {Wang}}, \ and\
  \bibinfo {author} {\bibfnamefont {Xie}\ \bibnamefont {Chen}},\ }\bibfield
  {title} {\enquote {\bibinfo {title} {Fracton models on general
  three-dimensional manifolds},}\ }\href@noop {} {\bibfield  {journal}
  {\bibinfo  {journal} {arXiv preprint arXiv:1712.05892}\ } (\bibinfo {year}
  {2017})}\BibitemShut {NoStop}%
\bibitem [{\citenamefont {Shi}\ and\ \citenamefont
  {Lu}(2017)}]{shi2017decipher}%
  \BibitemOpen
  \bibfield  {author} {\bibinfo {author} {\bibfnamefont {Bowen}\ \bibnamefont
  {Shi}}\ and\ \bibinfo {author} {\bibfnamefont {Yuan-Ming}\ \bibnamefont
  {Lu}},\ }\bibfield  {title} {\enquote {\bibinfo {title} {Decipher the
  nonlocal entanglement entropy of fracton topological orders},}\ }\href@noop
  {} {\bibfield  {journal} {\bibinfo  {journal} {arXiv preprint
  arXiv:1705.09300}\ } (\bibinfo {year} {2017})}\BibitemShut {NoStop}%
\bibitem [{\citenamefont {Ma}\ \emph {et~al.}(2017{\natexlab{a}})\citenamefont
  {Ma}, \citenamefont {Schmitz}, \citenamefont {Parameswaran}, \citenamefont
  {Hermele},\ and\ \citenamefont {Nandkishore}}]{ma2017topological}%
  \BibitemOpen
  \bibfield  {author} {\bibinfo {author} {\bibfnamefont {Han}\ \bibnamefont
  {Ma}}, \bibinfo {author} {\bibfnamefont {AT}~\bibnamefont {Schmitz}},
  \bibinfo {author} {\bibfnamefont {SA}~\bibnamefont {Parameswaran}}, \bibinfo
  {author} {\bibfnamefont {Michael}\ \bibnamefont {Hermele}}, \ and\ \bibinfo
  {author} {\bibfnamefont {Rahul~M}\ \bibnamefont {Nandkishore}},\ }\bibfield
  {title} {\enquote {\bibinfo {title} {Topological entanglement entropy of
  fracton stabilizer codes},}\ }\href@noop {} {\bibfield  {journal} {\bibinfo
  {journal} {arXiv preprint arXiv:1710.01744}\ } (\bibinfo {year}
  {2017}{\natexlab{a}})}\BibitemShut {NoStop}%
\bibitem [{\citenamefont {He}\ \emph {et~al.}(2017)\citenamefont {He},
  \citenamefont {Zheng}, \citenamefont {Bernevig},\ and\ \citenamefont
  {Regnault}}]{he2017entanglement}%
  \BibitemOpen
  \bibfield  {author} {\bibinfo {author} {\bibfnamefont {Huan}\ \bibnamefont
  {He}}, \bibinfo {author} {\bibfnamefont {Yunqin}\ \bibnamefont {Zheng}},
  \bibinfo {author} {\bibfnamefont {B~Andrei}\ \bibnamefont {Bernevig}}, \ and\
  \bibinfo {author} {\bibfnamefont {Nicolas}\ \bibnamefont {Regnault}},\
  }\bibfield  {title} {\enquote {\bibinfo {title} {Entanglement entropy from
  tensor network states for stabilizer codes},}\ }\href@noop {} {\bibfield
  {journal} {\bibinfo  {journal} {arXiv preprint arXiv:1710.04220}\ } (\bibinfo
  {year} {2017})}\BibitemShut {NoStop}%
\bibitem [{\citenamefont {Schmitz}\ \emph {et~al.}(2017)\citenamefont
  {Schmitz}, \citenamefont {Ma}, \citenamefont {Nandkishore},\ and\
  \citenamefont {Parameswaran}}]{schmitz2017recoverable}%
  \BibitemOpen
  \bibfield  {author} {\bibinfo {author} {\bibfnamefont {AT}~\bibnamefont
  {Schmitz}}, \bibinfo {author} {\bibfnamefont {Han}\ \bibnamefont {Ma}},
  \bibinfo {author} {\bibfnamefont {Rahul~M}\ \bibnamefont {Nandkishore}}, \
  and\ \bibinfo {author} {\bibfnamefont {SA}~\bibnamefont {Parameswaran}},\
  }\bibfield  {title} {\enquote {\bibinfo {title} {Recoverable information and
  emergent conservation laws in fracton stabilizer codes},}\ }\href@noop {}
  {\bibfield  {journal} {\bibinfo  {journal} {arXiv preprint arXiv:1712.02375}\
  } (\bibinfo {year} {2017})}\BibitemShut {NoStop}%
\bibitem [{\citenamefont {Vijay}\ and\ \citenamefont
  {Fu}(2017)}]{vijay2017generalization}%
  \BibitemOpen
  \bibfield  {author} {\bibinfo {author} {\bibfnamefont {Sagar}\ \bibnamefont
  {Vijay}}\ and\ \bibinfo {author} {\bibfnamefont {Liang}\ \bibnamefont {Fu}},\
  }\bibfield  {title} {\enquote {\bibinfo {title} {A generalization of
  non-abelian anyons in three dimensions},}\ }\href@noop {} {\bibfield
  {journal} {\bibinfo  {journal} {arXiv preprint arXiv:1706.07070}\ } (\bibinfo
  {year} {2017})}\BibitemShut {NoStop}%
\bibitem [{\citenamefont {Vijay}(2017)}]{vijay2017isotropic}%
  \BibitemOpen
  \bibfield  {author} {\bibinfo {author} {\bibfnamefont {Sagar}\ \bibnamefont
  {Vijay}},\ }\bibfield  {title} {\enquote {\bibinfo {title} {Isotropic layer
  construction and phase diagram for fracton topological phases},}\ }\href@noop
  {} {\bibfield  {journal} {\bibinfo  {journal} {arXiv preprint
  arXiv:1701.00762}\ } (\bibinfo {year} {2017})}\BibitemShut {NoStop}%
\bibitem [{\citenamefont {Ma}\ \emph {et~al.}(2017{\natexlab{b}})\citenamefont
  {Ma}, \citenamefont {Lake}, \citenamefont {Chen},\ and\ \citenamefont
  {Hermele}}]{ma2017fracton}%
  \BibitemOpen
  \bibfield  {author} {\bibinfo {author} {\bibfnamefont {Han}\ \bibnamefont
  {Ma}}, \bibinfo {author} {\bibfnamefont {Ethan}\ \bibnamefont {Lake}},
  \bibinfo {author} {\bibfnamefont {Xie}\ \bibnamefont {Chen}}, \ and\ \bibinfo
  {author} {\bibfnamefont {Michael}\ \bibnamefont {Hermele}},\ }\bibfield
  {title} {\enquote {\bibinfo {title} {Fracton topological order via coupled
  layers},}\ }\href@noop {} {\bibfield  {journal} {\bibinfo  {journal} {arXiv
  preprint arXiv:1701.00747}\ } (\bibinfo {year}
  {2017}{\natexlab{b}})}\BibitemShut {NoStop}%
\bibitem [{\citenamefont {Hal\'asz}\ \emph {et~al.}(2017)\citenamefont
  {Hal\'asz}, \citenamefont {Hsieh},\ and\ \citenamefont
  {Balents}}]{halasz2017fracton}%
  \BibitemOpen
  \bibfield  {author} {\bibinfo {author} {\bibfnamefont {G\'abor~B.}\
  \bibnamefont {Hal\'asz}}, \bibinfo {author} {\bibfnamefont {Timothy~H.}\
  \bibnamefont {Hsieh}}, \ and\ \bibinfo {author} {\bibfnamefont {Leon}\
  \bibnamefont {Balents}},\ }\href {https://arxiv.org/abs/1707.02308} {\enquote
  {\bibinfo {title} {Fracton topological phases from strongly coupled spin
  chains},}\ } (\bibinfo {year} {2017}),\ \Eprint
  {http://arxiv.org/abs/arXiv:1707.02308} {arXiv:1707.02308} \BibitemShut
  {NoStop}%
\bibitem [{\citenamefont
  {Pretko}(2016{\natexlab{a}})}]{pretko2016subdimensional}%
  \BibitemOpen
  \bibfield  {author} {\bibinfo {author} {\bibfnamefont {Michael}\ \bibnamefont
  {Pretko}},\ }\bibfield  {title} {\enquote {\bibinfo {title} {Subdimensional
  particle structure of higher rank u (1) spin liquids},}\ }\href@noop {}
  {\bibfield  {journal} {\bibinfo  {journal} {arXiv preprint arXiv:1604.05329}\
  } (\bibinfo {year} {2016}{\natexlab{a}})}\BibitemShut {NoStop}%
\bibitem [{\citenamefont {Pretko}(2016{\natexlab{b}})}]{pretko2016generalized}%
  \BibitemOpen
  \bibfield  {author} {\bibinfo {author} {\bibfnamefont {Michael}\ \bibnamefont
  {Pretko}},\ }\bibfield  {title} {\enquote {\bibinfo {title} {Generalized
  electromagnetism of subdimensional particles: A spin liquid story},}\
  }\href@noop {} {\bibfield  {journal} {\bibinfo  {journal} {arXiv preprint
  arXiv:1606.08857}\ } (\bibinfo {year} {2016}{\natexlab{b}})}\BibitemShut
  {NoStop}%
\bibitem [{\citenamefont {Xu}(2006{\natexlab{a}})}]{xu2006novel}%
  \BibitemOpen
  \bibfield  {author} {\bibinfo {author} {\bibfnamefont {Cenke}\ \bibnamefont
  {Xu}},\ }\bibfield  {title} {\enquote {\bibinfo {title} {Novel algebraic
  boson liquid phase with soft graviton excitations},}\ }\href@noop {}
  {\bibfield  {journal} {\bibinfo  {journal} {arXiv preprint cond-mat/0602443}\
  } (\bibinfo {year} {2006}{\natexlab{a}})}\BibitemShut {NoStop}%
\bibitem [{\citenamefont {Xu}(2006{\natexlab{b}})}]{xu2006gapless}%
  \BibitemOpen
  \bibfield  {author} {\bibinfo {author} {\bibfnamefont {Cenke}\ \bibnamefont
  {Xu}},\ }\bibfield  {title} {\enquote {\bibinfo {title} {Gapless bosonic
  excitation without symmetry breaking: An algebraic spin liquid with soft
  gravitons},}\ }\href@noop {} {\bibfield  {journal} {\bibinfo  {journal}
  {Physical Review B}\ }\textbf {\bibinfo {volume} {74}},\ \bibinfo {pages}
  {224433} (\bibinfo {year} {2006}{\natexlab{b}})}\BibitemShut {NoStop}%
\bibitem [{\citenamefont {Pankov}\ \emph {et~al.}(2007)\citenamefont {Pankov},
  \citenamefont {Moessner},\ and\ \citenamefont
  {Sondhi}}]{pankov2007resonating}%
  \BibitemOpen
  \bibfield  {author} {\bibinfo {author} {\bibfnamefont {S.}~\bibnamefont
  {Pankov}}, \bibinfo {author} {\bibfnamefont {R.}~\bibnamefont {Moessner}}, \
  and\ \bibinfo {author} {\bibfnamefont {S.~L.}\ \bibnamefont {Sondhi}},\
  }\bibfield  {title} {\enquote {\bibinfo {title} {Resonating singlet valence
  plaquettes},}\ }\href {\doibase 10.1103/PhysRevB.76.104436} {\bibfield
  {journal} {\bibinfo  {journal} {Phys. Rev. B}\ }\textbf {\bibinfo {volume}
  {76}},\ \bibinfo {pages} {104436} (\bibinfo {year} {2007})}\BibitemShut
  {NoStop}%
\bibitem [{\citenamefont {Xu}\ and\ \citenamefont
  {Wu}(2008)}]{xu2008resonating}%
  \BibitemOpen
  \bibfield  {author} {\bibinfo {author} {\bibfnamefont {Cenke}\ \bibnamefont
  {Xu}}\ and\ \bibinfo {author} {\bibfnamefont {Congjun}\ \bibnamefont {Wu}},\
  }\bibfield  {title} {\enquote {\bibinfo {title} {Resonating plaquette phases
  in large spin cold atom systems},}\ }\href@noop {} {\bibfield  {journal}
  {\bibinfo  {journal} {arXiv preprint arXiv:0801.0744}\ } (\bibinfo {year}
  {2008})}\BibitemShut {NoStop}%
\bibitem [{\citenamefont {Xu}\ and\ \citenamefont
  {Ho{\v{r}}ava}(2010)}]{xu2010emergent}%
  \BibitemOpen
  \bibfield  {author} {\bibinfo {author} {\bibfnamefont {Cenke}\ \bibnamefont
  {Xu}}\ and\ \bibinfo {author} {\bibfnamefont {Petr}\ \bibnamefont
  {Ho{\v{r}}ava}},\ }\bibfield  {title} {\enquote {\bibinfo {title} {Emergent
  gravity at a lifshitz point from a bose liquid on the lattice},}\ }\href@noop
  {} {\bibfield  {journal} {\bibinfo  {journal} {Physical Review D}\ }\textbf
  {\bibinfo {volume} {81}},\ \bibinfo {pages} {104033} (\bibinfo {year}
  {2010})}\BibitemShut {NoStop}%
\bibitem [{\citenamefont {Rasmussen}\ \emph {et~al.}(2016)\citenamefont
  {Rasmussen}, \citenamefont {You},\ and\ \citenamefont
  {Xu}}]{rasmussen2016stable}%
  \BibitemOpen
  \bibfield  {author} {\bibinfo {author} {\bibfnamefont {Alex}\ \bibnamefont
  {Rasmussen}}, \bibinfo {author} {\bibfnamefont {Yi-Zhuang}\ \bibnamefont
  {You}}, \ and\ \bibinfo {author} {\bibfnamefont {Cenke}\ \bibnamefont {Xu}},\
  }\bibfield  {title} {\enquote {\bibinfo {title} {Stable gapless bose liquid
  phases without any symmetry},}\ }\href@noop {} {\bibfield  {journal}
  {\bibinfo  {journal} {arXiv preprint arXiv:1601.08235}\ } (\bibinfo {year}
  {2016})}\BibitemShut {NoStop}%
\bibitem [{\citenamefont {Prem}\ \emph {et~al.}(2018)\citenamefont {Prem},
  \citenamefont {Pretko},\ and\ \citenamefont
  {Nandkishore}}]{prem2018emergent}%
  \BibitemOpen
  \bibfield  {author} {\bibinfo {author} {\bibfnamefont {Abhinav}\ \bibnamefont
  {Prem}}, \bibinfo {author} {\bibfnamefont {Michael}\ \bibnamefont {Pretko}},
  \ and\ \bibinfo {author} {\bibfnamefont {Rahul~M}\ \bibnamefont
  {Nandkishore}},\ }\bibfield  {title} {\enquote {\bibinfo {title} {Emergent
  phases of fractonic matter},}\ }\href@noop {} {\bibfield  {journal} {\bibinfo
   {journal} {Physical Review B}\ }\textbf {\bibinfo {volume} {97}},\ \bibinfo
  {pages} {085116} (\bibinfo {year} {2018})}\BibitemShut {NoStop}%
\bibitem [{\citenamefont {Pretko}(2017{\natexlab{a}})}]{pretko2017emergent}%
  \BibitemOpen
  \bibfield  {author} {\bibinfo {author} {\bibfnamefont {Michael}\ \bibnamefont
  {Pretko}},\ }\bibfield  {title} {\enquote {\bibinfo {title} {Emergent gravity
  of fractons: Mach's principle revisited},}\ }\href@noop {} {\bibfield
  {journal} {\bibinfo  {journal} {Physical Review D}\ }\textbf {\bibinfo
  {volume} {96}},\ \bibinfo {pages} {024051} (\bibinfo {year}
  {2017}{\natexlab{a}})}\BibitemShut {NoStop}%
\bibitem [{\citenamefont {Pretko}(2017{\natexlab{b}})}]{pretko2017finite}%
  \BibitemOpen
  \bibfield  {author} {\bibinfo {author} {\bibfnamefont {Michael}\ \bibnamefont
  {Pretko}},\ }\bibfield  {title} {\enquote {\bibinfo {title}
  {Finite-temperature screening of u (1) fractons},}\ }\href@noop {} {\bibfield
   {journal} {\bibinfo  {journal} {Physical Review B}\ }\textbf {\bibinfo
  {volume} {96}},\ \bibinfo {pages} {115102} (\bibinfo {year}
  {2017}{\natexlab{b}})}\BibitemShut {NoStop}%
\bibitem [{\citenamefont {Pretko}\ and\ \citenamefont
  {Radzihovsky}(2017)}]{pretko2017fracton}%
  \BibitemOpen
  \bibfield  {author} {\bibinfo {author} {\bibfnamefont {Michael}\ \bibnamefont
  {Pretko}}\ and\ \bibinfo {author} {\bibfnamefont {Leo}\ \bibnamefont
  {Radzihovsky}},\ }\bibfield  {title} {\enquote {\bibinfo {title}
  {Fracton-elasticity duality},}\ }\href@noop {} {\bibfield  {journal}
  {\bibinfo  {journal} {arXiv preprint arXiv:1711.11044}\ } (\bibinfo {year}
  {2017})}\BibitemShut {NoStop}%
\bibitem [{\citenamefont {Gromov}(2017)}]{gromov2017fractional}%
  \BibitemOpen
  \bibfield  {author} {\bibinfo {author} {\bibfnamefont {Andrey}\ \bibnamefont
  {Gromov}},\ }\bibfield  {title} {\enquote {\bibinfo {title} {Fractional
  topological elasticity and fracton order},}\ }\href@noop {} {\bibfield
  {journal} {\bibinfo  {journal} {arXiv preprint arXiv:1712.06600}\ } (\bibinfo
  {year} {2017})}\BibitemShut {NoStop}%
\bibitem [{\citenamefont {Fradkin}\ and\ \citenamefont
  {Shenker}(1979)}]{fradkin1979phase}%
  \BibitemOpen
  \bibfield  {author} {\bibinfo {author} {\bibfnamefont {Eduardo}\ \bibnamefont
  {Fradkin}}\ and\ \bibinfo {author} {\bibfnamefont {Stephen~H.}\ \bibnamefont
  {Shenker}},\ }\bibfield  {title} {\enquote {\bibinfo {title} {Phase diagrams
  of lattice gauge theories with higgs fields},}\ }\href {\doibase
  10.1103/PhysRevD.19.3682} {\bibfield  {journal} {\bibinfo  {journal} {Phys.
  Rev. D}\ }\textbf {\bibinfo {volume} {19}},\ \bibinfo {pages} {3682--3697}
  (\bibinfo {year} {1979})}\BibitemShut {NoStop}%
\bibitem [{Note1()}]{Note1}%
  \BibitemOpen
  \bibinfo {note} {The ground state degeneracy on a 3-torus of the exactly
  solvable X-cube model is stable under local perturbation. This can be
  verified using degenerate perturbation theory. If we add a local perturbation
  $V= \lambda \DOTSB \tsum \slimits@ \protect \mathcal {O}_{loc}$ to the
  Hamiltonian, the matrix elements in the resulting effective Hamiltonian for
  the degenerate ground state space are proportional to $\protect \mathaccentV
  {tilde}07E{\lambda }^L$, where $\protect \mathaccentV {tilde}07E{\lambda }
  \propto \lambda $ is a constant, and $L$ is the linear size of the system.
  This holds because only logical operators supported on a region of size $ L$
  or larger have non-vanishing matrix elements within the ground state
  subspace. Since the degenerate subspace has dimension $\sim c^L$ for some
  constant $c$, the matrix Frobenius norm is bounded by $(c \protect
  \mathaccentV {tilde}07E{\lambda })^L$, which in turn bounds all the
  eigenvalues. Therefore, as long as $\lambda $ is below some finite threshold,
  the splitting of the ground state subspace is exponentially small and
  approaches to zero in the thermodynamic limit. A very similar argument
  applies to many other gapped fracton models.}\BibitemShut {Stop}%
\bibitem [{\citenamefont {{Nandkishore}}\ and\ \citenamefont
  {{Hermele}}(2018)}]{fractons_review}%
  \BibitemOpen
  \bibfield  {author} {\bibinfo {author} {\bibfnamefont {R.~M.}\ \bibnamefont
  {{Nandkishore}}}\ and\ \bibinfo {author} {\bibfnamefont {M.}~\bibnamefont
  {{Hermele}}},\ }\bibfield  {title} {\enquote {\bibinfo {title}
  {{Fractons}},}\ }\href@noop {} {\bibfield  {journal} {\bibinfo  {journal}
  {ArXiv e-prints}\ } (\bibinfo {year} {2018})},\ \Eprint
  {http://arxiv.org/abs/1803.11196} {arXiv:1803.11196 [cond-mat.str-el]}
  \BibitemShut {NoStop}%
\bibitem [{\citenamefont {Kitaev}(2003)}]{kitaev2003fault}%
  \BibitemOpen
  \bibfield  {author} {\bibinfo {author} {\bibfnamefont {A~Yu}\ \bibnamefont
  {Kitaev}},\ }\bibfield  {title} {\enquote {\bibinfo {title} {Fault-tolerant
  quantum computation by anyons},}\ }\href@noop {} {\bibfield  {journal}
  {\bibinfo  {journal} {Annals of Physics}\ }\textbf {\bibinfo {volume}
  {303}},\ \bibinfo {pages} {2--30} (\bibinfo {year} {2003})}\BibitemShut
  {NoStop}%
\bibitem [{\citenamefont {Bulmash}\ and\ \citenamefont
  {Barkeshli}(2018)}]{bulmash2018higgs}%
  \BibitemOpen
  \bibfield  {author} {\bibinfo {author} {\bibfnamefont {Daniel}\ \bibnamefont
  {Bulmash}}\ and\ \bibinfo {author} {\bibfnamefont {Maissam}\ \bibnamefont
  {Barkeshli}},\ }\bibfield  {title} {\enquote {\bibinfo {title} {The higgs
  mechanism in higher-rank symmetric $ u (1) $ gauge theories},}\ }\href@noop
  {} {\bibfield  {journal} {\bibinfo  {journal} {arXiv preprint
  arXiv:1802.10099}\ } (\bibinfo {year} {2018})}\BibitemShut {NoStop}%
\end{thebibliography}%

\appendix
\widetext

\section{Ground state degeneracy and logical operators of the $\mathbb{Z}_2$ scalar charge theory\label{app:degeneracy}}

In this section, we calculate the ground state degeneracy in the deconfined phase of the rank-2 $\mathbb{Z}_2$ gauge theory on the cubic lattice.  We also discuss the logical operators and argue that the degenerate ground states are locally indistinguishable, \emph{i.e.} the ground state degeneracy is topological in nature.  We work at the exactly solvable point and use the description in terms of a bosonic model (\emph{i.e.} one with a tensor product Hilbert space), whose Hamiltonian is given by Eq. (\ref{eq:ham_rank-2_z2-bosonic}) with $U = J = 0$.

In a $L\times L \times L$ system, there are $L^3$ unit cells with four distinct sites within each unit cell. Three sites lie on the plaquettes, with one qubit on each plaquette. The other site lies on the vertex and has three qubits. In total, the number of qubits is $6L^3$.  We always take $L$ to be even.

We count the number of independent stabilizers. There is one octahedron term $G_{\bf r}$ per vertex, thus we have $L^3$ of these terms in total.  However, these terms are not all independent and satisfy some constraints.  We separate the vertices into eight groups, according to the parity $p_\mu = r_{\mu} \mod 2$ of the coordinates ($\mu = x,y,z$), labeling the groups by $(p_x, p_y, p_z)$.  The product $G_{(p_x, p_y, p_z)} = \prod_{\bf r \in (p_x, p_y, p_z)} G_{\bf r}$ is a product of $X_{xy}$ over $xy$ planes with $z \mod 2 = p_z$, $X_{yz}$ over $yz$ planes with $x \mod 2 = p_x$, and $X_{zx}$ over $zx$ planes with $y \mod 2 = p_y$. We can think of the operators $G_{(p_x, p_y, p_z)}$ as lying at the vertices of a cube, and it is easy to check that the product of these operators over any face of the cube is unity.  This gives four independent constraints, and the number of independent vertex terms is $L^3-4$.

The magnetic field terms are more complicated, and to count them we resort to a numerical method employed previously in Ref.~\onlinecite{ma2017fracton}, which we also use to check the counting of $X$ stabilizers above.  A product of $X$ or $Z$ Pauli operators is viewed as an element of the ${\mathbb F}_2$ vector space $V_{X(Z)} \simeq ({\mathbb F}_2)^{6 L^3}$, with vector addition corresponding to operator multiplication.  The set of all $X$ ($Z$) stabiliers is a subspace $S_{X (Z)} \subset V_{X(Z)}$, and the number of independent stabilizers is the dimension of this subspace.  Any $X$ stabilizer is a product of the $L^3$ operators $G_{\bf r}$; in linear algebra language, we can say that $G_{\bf r}$ gives a spanning set for $S_X$.  This spanning set can be represented as a $L^3 \times 6 L^3$ matrix, and its rank, which can be determined via row reduction, gives the dimension of $S_X$.  The same method can be applied to $Z$ stabilizers, where $F_{\mu \nu}$ constitute a generating set.  We used this method for even $L = 2,\dots,12$ to confirm that the number of independent $X$-stabilizers is $L^3 - 4$, and to determine that the number of independent $Z$-stabilizers is $5L^3 - 8$.

This implies that the ground state degeneracy is $\log_2 (GSD)=6L^3- (L^3-4)-(5L^3 - 8) = 12$, which is equal to the degeneracy of four copies of the toric code on a 3-torus.  To check that the degeneracy is topological in nature, using the same numerical method as above, we return to the space $S_Z$ of $Z$-stabilizers described by its generating set.  For each type of charge $\tau^z_{1},\dots, \tau^z_4$, we add to the generating set three large string operators that transport the charge around the three cycles of the 3-torus.  These string operators are products of $Z_{\mu \mu}$, and we choose them to run along straight lines, with their transverse position arbitrary.  Upon adding these operators to the generating set, we now find $5 L^3 + 4$ independent $Z$-operators, including both $Z$-stabilizers and string logical operators.  Adding the string operators thus fully resolves the space of degenerate ground states.  Moreover, because the transverse position of the string operators is arbitrary, they can be chosen to avoid the position of any local operator of interest, and it follows that the ground states are locally indistinguishable, as expected for four copies of the $d=3$ toric code.

\section{$\mathbb{Z}_2$ scalar charge theory in two dimensions \label{app:rank-2_Z2_2d}}

\begin{figure}[h]
\includegraphics[width=.6\textwidth]{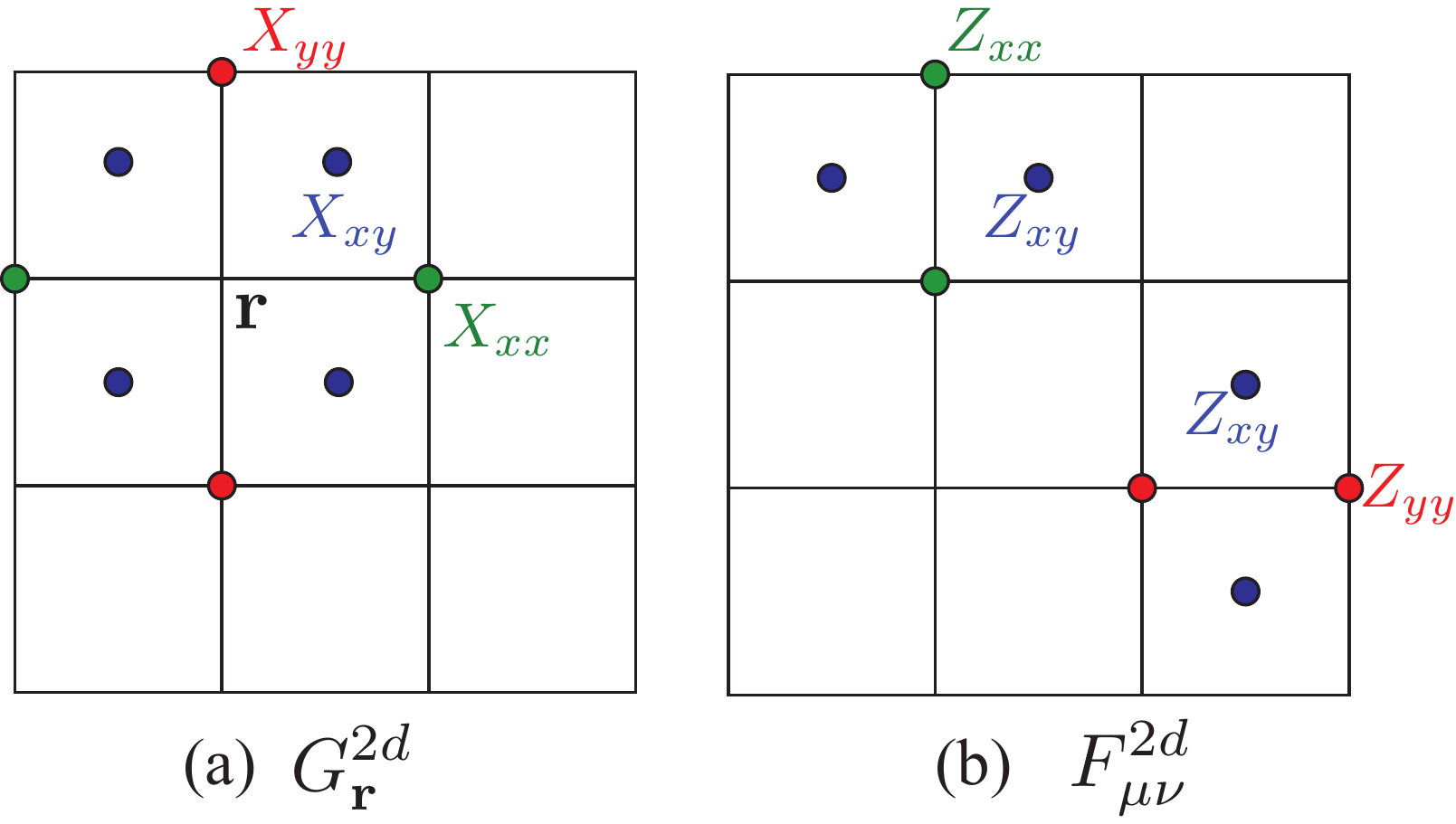}
\caption{(a)Plaquette terms generated by the Gauss law. (b) Gauge invariant terms. \label{fig:scalar_terms}}
\end{figure}

In this section, we construct the rank-2 $\mathbb{Z}_2$ scalar charge theory on the square lattice in two dimensions. The $\mathbb{Z}_2$ variables are defined in a similar way as in $d=3$: each site has two diagonal components while the off-diagonal components are defined on the plaquettes. The Hamiltonian is
\begin{equation}
H_{\mathbb{Z}_2}^{2d} = -K \sum_{{\bf r},\mu,\nu} F^{2d}_{\mu\nu} - u \sum_{\bf r} G^{2d}_{\bf r}
\end{equation}
where $F_{\mu\nu}$ is a product of 4 $Z_{\mu\nu}$ operators around each link and $ G_{\bf r}$ is a product of 8 $X_{\mu\nu}$ operators around a site, as shown in Fig.~\ref{fig:scalar_terms}. 

We now show that this model is equivalent to three copies of the $d=2$ toric code. Numerically, using the method we reviewed in Appendix~\ref{app:degeneracy}, we obtained the ground state degeneracy of this model on a $L \times L$ 2-torus, with even $L=2,\dots,20$, finding $\log_2 ({\rm GSD})= 6$. This is the expected result for three copies of the $d=2$ toric code.

\begin{figure}[h]
\includegraphics[width=.9\textwidth]{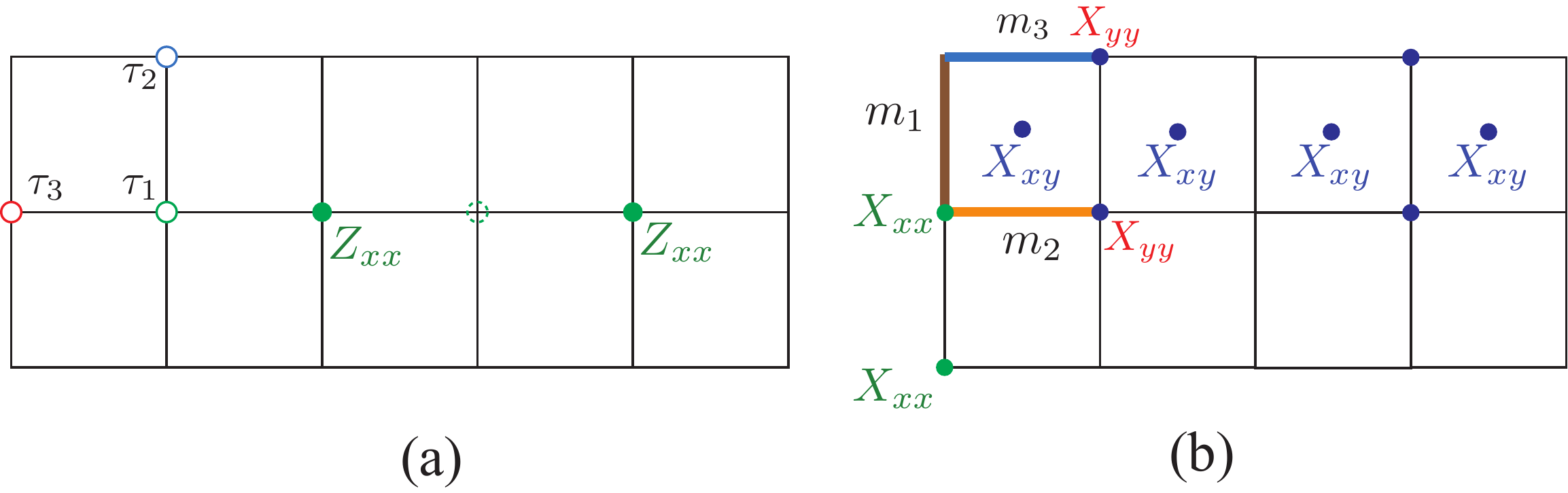}
\caption{(a) Three distinct charges in the rank-2 scalar charge theory on the square lattice. A string operator for $\tau_1$ is shown. (b) Three disinct fluxes $m_1, m_2, m_3$. We also show a string operator along $y$ direction and a thick string operator along $x$ direction for $m_1$. \label{fig:em}}
\end{figure}

Point-like charge excitations are created at the sites by acting with $Z_{\mu\nu}$. Acting with $Z_{\mu\mu}({\bf r})$, we create two charge excitations at the two neighbors of ${\bf r}$ in the $\pm \mu$ direction. Charges can thus hop two lattice spacings in the $x$ and $y$ directions (Fig.~\ref{fig:em}a). Naively, it might seem that the four charges at corners of a plaquette are distinct types of excitations.  But since these four charges can be created together by acting with $Z_{xy}$, there are actually three different types of charges.

Acting  with $X_{\mu\nu}$  creates magnetic flux excitations living on the links, denoted as $m_{1,2,3}$ (Fig.~\ref{fig:em}(b)). They are also mobile excitations. For example, $m_1$ can be transported along the $y$ direction by acting with a string operator that is a product of $X_{xx}$ operators along the $y$ direction. The same excitation can be transported along the $x$ direction by applying a ``thick'' string operator, as shown in Fig.~\ref{fig:em}(b). $m_1$ can hop one lattice spacing in the $y$-direction, and two lattice spacings in the $x$-direction. The other flux excitations have similar mobility. Similar to the charge excitations, four flux excitations on the edges of a plaquette are created together by acting with $X_{xy}$, which means that there are three rather than four distinct types of fluxes.

Next, we can check the braiding between magnetic fluxes and electric charges by simply considering the commutation relation between the corresponding string operators at a crossing point. The result is listed in Table~\ref{tab:braiding_2d}. It is straightforward to verify that the flux type is fully resolved by braiding with the three different electric charges.

\begin{table}
\begin{tabular}{c|c}
Electric charges & Magnetic fluxes with $\theta = \pi$ statistics \\
\hline
$\tau_1$ &  $m_1$, $m_3$  \\
\hline
$\tau_2$ &  $m_1$, $m_2$  \\
\hline
$\tau_3$ &  $m_3$  \\
\hline
\end{tabular}
\caption{In this table, for each type of electric charges as shown in Fig.~\ref{fig:em}(a), we list the magnetic fluxes $m_{1,2,3}$ that acquire a statistical phase $\theta = \pi$ when braided around the electric charge.  The statistical phase is $\theta = 0$ for charge-flux pairs not listed.
\label{tab:braiding_2d}}
\end{table}

All the analysis above, including the ground state degeneracy, logical operators and properties of the excitations, shows that the rank-2 $\mathbb{Z}_2$ scalar charge theory is three copies of $d=2$ toric code, which is of course a conventional topological order without any fracton or sub-dimensional excitations. This is consistent with our prediction from Higgsing the conservation law in Sec.~\ref{sec:dipole}. In three dimensions, we can gap the diagonal components of the tensor fields to get a hollow tensor gauge theory describing fracton topological order. However, in two dimensions, the corresponding hollow gauge theory gives a classical state, because it lacks a gauge-invariant flux term.

\end{document}